\documentclass[aps,prx,reprint,showpacs,amsfonts,amsmath,floatfix,superscriptaddress,footinbib]{revtex4-2}

\usepackage[svgnames]{xcolor}

 \usepackage[colorlinks=true,        
            allcolors = black,  
            citecolor=DarkBlue, 
            linkcolor=DarkBlue,
            urlcolor=DarkBlue
        ]{hyperref}  
\usepackage{amsmath, amsfonts, amssymb, amsthm, bbm}
\usepackage{dsfont}
\usepackage{bm} 
\usepackage{mathtools}
\usepackage{cprotect}
\usepackage{times}
\usepackage{multirow}
\usepackage{physics}
\usepackage{graphicx}
\usepackage{float}
\usepackage{enumitem}
\usepackage{booktabs}
\usepackage{dcolumn}
\usepackage{diagbox}
\usepackage{qcircuit}
\usepackage[english, german]{babel}
\usepackage{orcidlink}

\usepackage{letltxmacro}
\usepackage{comment}

\LetLtxMacro{\ORIGselectlanguage}{\selectlanguage}
\makeatletter
\DeclareRobustCommand{\selectlanguage}[1]{%
  \@ifundefined{alias@\string#1}
    {\ORIGselectlanguage{#1}}
    {\begingroup\edef\x{\endgroup
       \noexpand\ORIGselectlanguage{\@nameuse{alias@#1}}}\x}%
}
\newcommand{\definelanguagealias}[2]{%
  \@namedef{alias@#1}{#2}%
}
\makeatother

\definelanguagealias{en}{english}

\newcommand{\Aop}[1]{A_{#1}} 
\newcommand{\acoeff}[2]{a_{#2}({#1})} 
\newcommand{\bcoeff}[2]{b_{#2}({#1})}
\newcommand{\ccoeff}[2]{c_{#2}({#1})}
\newcommand{\hterm}[1]{h_{#1}}  
\newcommand{\Psistep}[1]{\Psi_{#1}} 
\newcommand{\psuccess}{p_{GS}}

\allowdisplaybreaks
\begin{document}
\selectlanguage{english}
\title{Constant-Depth Quantum Imaginary Time Evolution Using Dynamic Fan-out Circuits}

\author{Albert Lund \orcidlink{0009-0008-8374-529X}}
\email{lundalb@chalmers.se}
\affiliation{Jeppesen ForeFlight, Gothenburg 411 03, Sweden}
\affiliation{Department of Microtechnology and Nanoscience (MC2), Chalmers University of Technology, 412 96 Gothenburg, Sweden}

\author{Erika Magnusson \orcidlink{0009-0002-1948-487X}}
\affiliation{Department of Chemistry and Chemical Engineering, Chalmers University of Technology, 412 96 Gothenburg, Sweden}

\author{Werner Dobrautz \orcidlink{0000-0001-6479-1874}}
\affiliation{Center for Advanced Systems Understanding, Helmholtz-Zentrum Dresden-Rossendorf, Germany}
\affiliation{Center for Scalable Data Analytics and Artificial Intelligence Dresden/Leipzig, TU Dresden, Germany}
 
\author{Laura Garc\'{\i}a-\'{A}lvarez \orcidlink{0000-0002-3367-8083}}
\affiliation{Department of Microtechnology and Nanoscience (MC2), Chalmers University of Technology, 412 96 Gothenburg, Sweden}


\begin{abstract}

Dynamic quantum circuits combine mid-circuit measurement with classical feed-forward, enabling circuit constructions with reduced entangling-gate depth.
Here, we investigate their use in Quantum Imaginary Time Evolution (QITE), where circuit depth and parameter growth limit practical implementations of ground-state preparation.
For dense classical optimization Hamiltonians, we introduce a reduced-parameter QITE ansatz that restricts entanglement generation via a small set of control qubits, enabling each QITE layer to be implemented with constant two-qubit gate depth using fan-out-based dynamic circuits.
In noiseless simulations of exact cover and set partitioning instances, the reduced ansatz yields a higher success probability than standard QITE approaches.
We implement unitary, dynamic-fan-out, and semi-classical adaptive variants on IBM superconducting hardware.
The semi-classical variant performs favorably to the unitary implementation, while the fully dynamic construction exposes the trade-offs between entangling-depth reduction and measurement and feed-forward overhead associated to dynamic circuit implementations.
Using a fidelity threshold of 0.5 relative to the noiseless QITE ansatz, we show that dynamic fan-out based QITE would outperform unitary implementations on current devices when the measurement and two-qubit gate errors are reduced by $65\%$ and the feedback latency is halved.
\end{abstract}

\maketitle

\section{Introduction}
Dynamic quantum circuits integrate mid‑circuit measurements with classical feed‑forward to condition quantum operations on measurement outcomes.
Such adaptivity is central in a range of quantum information protocols, including for example: Measurement-based quantum computation, which relies on sequences of single‑qubit measurements whose bases depend on earlier outcomes~\cite{briegel2009measurement-based}. 
Teleportation-gate schemes, where local operations and classical communication (LOCC) enable efficient circuit cutting/knitting techniques for distributed quantum computing~\cite{piveteau2024circuit,carrera-vazquez2024combining}.
Fault-tolerant quantum error correction, in which stabilizer measurements will need to be processed in real time to guide corrective operations~\cite{Bluvstein_2023, Sivak_2023, Acharya_Suppressing_2023, Bluvstein_Geim_Li_Evered_Bonilla_Ataides_Baranes_Gu_Manovitz_Xu_Kalinowski}.

Adaptivity functions not only as a practical tool; its computational power has also been explored within several different theoretical models. 
One line of work is Local Alternating Quantum–Classical Computation (LAQCC) circuits, which provide a unifying framework for describing adaptive operations~\cite{buhrman_state_2024}, including LOCC-assisted circuits~\cite{Piroli_2021}.
Although the computational power of these so-called dynamic circuits does not exceed that of logarithmic‑depth unitary circuits, they can yield practical circuit depth reductions in current and near-future devices prior to full fault‑tolerance. 
More concretely, they enable quantum state preparation in reduced circuit depth~\cite{Piroli_2021, 2022_tsungcheng_mesurement, Piroli_2024, Yeo_2025_reducing}, including constant-time constructions~\cite{buhrman_state_2024, Smith_2024_constant, sahay2024finitedepthpreparationtensornetwork, zi2025constantdepthquantumcircuitsarbitrary, Iqbal_2024, baumer_measurement-based_2024, song_constant_2025,baumer_2024_efficient}.
LAQCC circuits also contain important classes such as Instantaneous Quantum Polynomial‑Time circuits ~\cite{buhrman_state_2024,cao2025measurementdrivenquantumadvantagesshallow}, which are believed to be classically intractable \cite{Bremner_2010}.
Another formal study characterizes oracle separation of hybrid quantum–classical models ~\cite{arora2022oracle}. 
The work shows that adaptive shallow‑depth quantum circuits are incomparable to models that repeatedly query a shallow quantum circuit from a classical computer, as done in variational algorithms: each model can efficiently solve oracle problems the other computational model cannot. 
Combining these features in adaptive shallow-depth quantum circuits that can be queried in a classical computation yields another hybrid model that is strictly more powerful than either of the previous ones alone; however, it still remains below full polynomial‑depth quantum computation. 
A third perspective comes from circuit families such as Clifford circuits, where conditioning gates on measurement outcomes breaks classical simulability. 
That is, adaptive Clifford circuits cannot be efficiently classically simulated, unlike their non-adaptive counterparts~\cite{jozsa2014classical}.

The community is now focused on probing adaptivity and testing its performance on current hardware.
Recent experiments demonstrate constant-depth implementations of fan-out operations, CNOT ladders, and related primitives~\cite{baumer_measurement-based_2024, song_constant_2025}, showcasing a fidelity increase of $\gtrsim 0.1$ for a non-local CNOT gate for distances larger than 20 qubits on devices with restricted connectivity~\cite{baumer_2024_efficient}.
Dynamic circuits have also enabled the execution of an 80-qubit circuit on a 20-qubit trapped‑ion processor through qubit‑reuse compilation~\cite{decross2023qubit-reuse}.
These developments highlight the potential of dynamic circuits for algorithms where non-local interactions dominate circuit depth. 
Mid-circuit measurements and feed‑forward can compress key primitives and inspire new heuristics tailored to today's hardware. 
Moreover, recent barren‑plateau‑free methods~\cite{Yan_2025, deshpande2025dynamicparameterizedquantumcircuits} and conjectured quantum advantage schemes~\cite{huang2025generativequantumadvantageclassical, Watts_2025_quantum, cao2025measurementdrivenquantumadvantagesshallow} already incorporate dynamic‑circuit elements.

This perspective naturally connects to algorithms such as Quantum Imaginary Time Evolution (QITE), where depth‑efficient constructions are essential for near-term use.
QITE is a ground-state preparation method that approximates normalized imaginary time evolution (ITE) using unitary operations~\cite{motta_determining_2020}. 
While imaginary time evolution converges exponentially to the ground state in principle, its implementation on quantum hardware is challenging due to its non-unitary nature. 
Practical quantum algorithms therefore rely on unitary approximations (in the case of QITE and similar) or variational formulations (such as Variational Quantum Imaginary Time Evolution, VarQITE \cite{McArdle_2019}), which typically require non-local operators or are restricted to parametrized ansätze respectively.
As a result of the unitary approximation, the circuit depth in QITE grows with system size and the number of imaginary-time steps~\cite{motta_determining_2020,Nishi_Kosugi_Matsushita_2021,Gomes_Zhang_Berthusen_Wang_Ho_Orth_Yao_2020}.  
Thus, we raise a practical question: under what conditions can dynamic circuits reduce the entangling-gate depth of QITE, and when do the additional measurement and feed-forward operations outweigh these gains on current quantum hardware?

As a concrete testbed, we study QITE applied to classical constraint satisfaction and optimization problems, focusing on exact cover and set partitioning instances derived from the tail assignment problem~\cite{Grönkvist_2005, Vikstal_2020_applying, Willsch_2022, Svensson_2023_Hybrid}. 
These problems map to diagonal Hamiltonians with pairwise interactions and admit instances with dense conflict graphs, where many variables participate in overlapping constraints. 
Such dense instances exacerbate the non-local structure of QITE operators and therefore provide a suitable setting to investigate depth-compression strategies based on dynamic circuits.

In this work, we introduce a reduced-parameter QITE ansatz tailored to dense classical Hamiltonians, which restricts entanglement generation to a small set of control qubits. 
This structure allows each QITE layer to be implemented with constant two-qubit gate depth using fan-out-based dynamic circuits. 
We validate the resulting ansatz in noiseless simulations on exact cover and set partitioning instances, and compare unitary, dynamic-fan-out, and semi-classical adaptive implementations on IBM superconducting hardware. 
By combining experimental results with a fidelity and timing analysis, we identify the regimes in which dynamic circuit constructions are expected to outperform purely unitary implementations.

The paper is structured as follows. Section~\ref{sec:preliminaries} reviews QITE, previously proposed adjustments, and introduces the set partitioning and exact cover problem. 
Section~\ref{sec:methods} presents the reduced-parameter QITE ansatz and outlines unitary, dynamic and semi-classical hardware implementations. Section~\ref{sec:results} reports simulation and hardware results. Section~\ref{sec:conclusion} concludes with a discussion and outlook.

\section{Preliminaries}
\label{sec:preliminaries}
This section reviews QITE and introduces the classical optimization problems used in this work.

\subsection{Quantum imaginary time evolution}
\label{sec:prelim_QITE}
ITE is a projector method based on a Wick rotation of the Schrödinger equation, in which the time $t$ is replaced as $t \mapsto -i\tau$, leading to evolution under the non-unitary operator $e^{-\tau H}$~\cite{Wick_1954}.
Given an initial state with non-zero overlap with the ground state, ITE suppresses excited-state components exponentially in $\tau$, yielding convergence to the ground state in the limit $\tau \to \infty$.

Implementing ITE on a quantum computer is challenging due to its non-unitary nature. 
Several approaches have been proposed to address this difficulty. 
Block-encoding based methods implement non-unitary evolution probabilistically and typically require amplitude amplification to boost success probability, making them more suitable in fault-tolerant regimes~\cite{nishi2022accelerationprobabilisticimaginarytimeevolution, Silva_2023, Kosugi_2022_imaginary, leadbeater2023nonunitarytrottercircuitsimaginary, Xie_probabilistic_2024}. 
Double-bracket constructions provide alternative iterative schemes but incur gate counts that scale exponentially in the number of iterations~\cite{gluza2025doublebracketquantumalgorithmsquantum}. 
Variational formulations, including VarQITE~\cite{McArdle_2019, Kolotouros_2025, Benedetti_2021_hardware, wang2023symmetryenhancedvariationalquantum, Gacon_Variational_2024, Sokolov2023, Magnusson2024, Dobrautz2024}, restrict the evolution to a parameterized ansatz. 
Hybrid and machine-learning inspired approaches have also been explored~\cite{Rrapaj_2025, zhong2024classicaloptimizationimaginarytime, kuji2025variationalquantumneuralhybridimaginary, xie2025adaptiveweightedqitevqealgorithm, chai2025optimizingquboquantumcomputer}. 

In this work, we focus on the tomography-based QITE algorithm of Motta \textit{et al.}~\cite{motta_determining_2020}. 
There, the non-unitary operator $e^{-\Delta\tau \hterm{m}}$ associated with a geometrically local Hamiltonian $H = \sum_m \hterm{m}$ applied to a state $\ket{\Psistep{m-1}}$ is approximated at each step by a parameterized unitary operator $e^{-i\Delta\tau \Aop{m}}$ acting on a region of support of $D$ qubits around $\hterm{m}$,
\begin{equation}
    e^{-i\Delta\tau \Aop{m}} \ket{\Psistep{m-1}} 
    \approx 
    \frac{e^{-\Delta\tau \hterm{m}} \ket{\Psistep{m-1}}}
    {\| e^{-\Delta\tau \hterm{m}} \ket{\Psistep{m-1}} \|}.
\end{equation}
$\Aop{m}$ is expanded in some operator basis
\begin{equation}
    \label{eq:a_operator}
    \Aop{m} = \sum_I \acoeff{m}{I} \sigma_I ,
\end{equation}
where $\{\sigma_I\}$ denotes a chosen operator basis, typically a subset of the $N$-qubit Pauli group.
The coefficients $\acoeff{m}{I}$ are obtained by solving a linear system $\mathbf{S} \acoeff{m}{} = \mathbf{b}$. 
For a self-adjoint operator basis, the matrix $\mathbf{S}$ and vector $\mathbf{b}$ are given by
\begin{align}
\label{eq:qite_setup}
    &S_{IJ} = \langle \Psistep{m-1} | \sigma_I^\dagger \sigma_J | \Psistep{m-1} \rangle , \nonumber \\
    &\mathbf{b}_I = \frac{i}{\sqrt{c}} \langle \Psistep{m-1} | [\hterm{m}, \sigma_I] | \Psistep{m-1} \rangle , \nonumber \\
    &c = 1 - 2\Delta\tau \langle \Psistep{m-1} | \hterm{m} | \Psistep{m-1} \rangle ,
\end{align}
where $c$ is the first-order approximation of the normalization factor 
$\langle \Psistep{m-1} | e^{-2\Delta\tau \hterm{m}} | \Psistep{m-1} \rangle$.
This construction implements imaginary time evolution up to an error of  $\mathcal{O}(\Delta\tau^2)$ per imaginary-time step.

The quantum resources required by QITE depend on the size of the operator basis used to expand $\Aop{m}$. 
If all Pauli strings supported on $D$ qubits are included, the expansion of $\Aop{m}$ requires $O(4^D)$ parameters, and the construction of the corresponding linear system requires a number of measurements and classical storage that scale exponentially in $D$~\cite{motta_determining_2020}. 
For Hamiltonians with finite interaction range and sufficiently geometrically local structure, restricting the operator support to small $D$ can provide accurate approximations in practice. 
However, for Hamiltonians with long-range interactions, small-support approximations may fail to capture the relevant correlations. 
Enlarging the operator basis can significantly improve convergence, though at the expense of increased measurement cost, classical processing, and circuit depth~\cite{Nishi_Kosugi_Matsushita_2021, Gomes_Zhang_Berthusen_Wang_Ho_Orth_Yao_2020}.

\subsection{Exact Cover and Set Partitioning as Diagonal Hamiltonians}
\label{sec:exact_cover}

We now specialize QITE to Hamiltonians that are diagonal in the computational basis, as arise naturally in classical binary optimization problems.
Such Hamiltonians consist of a sum of commuting interactions that include Pauli $Z$ operators and their tensor products, and can therefore be treated without Trotterization over individual Hamiltonian terms.
Moreover, when starting from a state with real amplitudes in the computational basis, the QITE linear system simplifies substantially: as shown in Ref.~\cite{motta_determining_2020}, the vector $\mathbf{b}$ vanishes for Pauli strings containing an even number of $Y$ components.
Consequently, only Pauli strings with an odd number of $Y$ operators can contribute to the generator $\Aop{m}$ at each imaginary-time step.

As a concrete testbed within this problem class, we consider the exact cover problem and its optimization variant, set partitioning.
Given a collection $R$ of subsets of a set $F$, an exact cover is a subcollection $R^* \subseteq R$ such that each element $f \in F$ is contained in exactly one subset of $R^*$.
Exact cover arises naturally in the tail assignment problem, where a set of flights $F$ must be covered by a subset of feasible routes $R$, often generated from column-generation subproblems representing possible flight routes of an aircraft~\cite{Grönkvist_2005}.
Beyond tail assignment, exact cover appear in applications such as web service composition~\cite{Ba_2016}, polyomino puzzles~\cite{Nishino_Yasuda_Minato_Nagata_2017}, and stock cutting~\cite{Balas_Padberg_1976}.
Previous quantum algorithm studies of exact cover have employed the Quantum Approximate Optimization Algorithm (QAOA)~\cite{Vikstal_2020_applying, Svensson_2023_Hybrid, saavedrapino2025quantumapproximateoptimizationalgorithm, ni2025quantumassistedrecursivealgorithmsolving}, quantum annealing~\cite{Willsch_2022}, the Variational Quantum Eigensolver (VQE)~\cite{cacao_2024_the}, and evolutionary quantum algorithms~\cite{fernandez2025quantumcircuitevolutionaryframework}.

The set partitioning problem extends exact cover by associating a cost $c_r$ with each subset $r$ representing a route and minimizing the total cost of the selected subsets $x_r$ subject to the exact cover constraints.
Following Refs.~\cite{Vikstal_2020_applying, Svensson_2023_Hybrid}, we formulate set partitioning as a simplified tail assignment problem~\cite{Grönkvist_2005},
\begin{eqnarray}
     &&\min_{x} \sum_{r \in R} c_r x_r, \nonumber \\
     &&\text{subject to } \sum_{r \in R} a_{fr}x_r = 1 \quad \forall f \in F, \nonumber \\
     && x_r \in \{0,1\} \quad \forall r \in R,
     \label{eq:set_partitioning}
\end{eqnarray}
where $F$ denotes the set of flights to be covered and $R$ the set of legal routes. 
The coefficients $a_{fr}$ define the available routes such that $a_{fr}=1$ if flight $f$ is contained in route $r$ and otherwise $0$.
Discarding the objective function yields the exact cover constraint satisfaction problem.
In this sense, exact cover is a feasibility problem where it is sufficient to find an assignment satisfying the constraints whereas set partitioning is an optimization problem seeking the optimal assignment.

The Ising Hamiltonian corresponding to set partitioning is
\begin{equation}
    \label{eq:exact_cover_Hamiltonian}
    H = \sum_{r \in R} \big(\mu_r h_r + h'_r\big) Z_r
        + \sum_{r < r'} \mu_r J_{rr'} Z_r Z_{r'},
\end{equation}
with
\begin{eqnarray}
&&h'_r = \frac{c_r}{2},\\
&&h_r = \sum_{f \in F} a_{fr} \Big( \sum_{r' \in R} \frac{a_{fr'}}{2} - 1 \Big), \\
&&J_{rr'} = \sum_{f \in F} \frac{a_{fr} a_{fr'}}{2},
\end{eqnarray}
where $\mu_r$ are penalty parameters enforcing the constraints.
We define an instance as dense if it has a large proportion of non-zero $J_{rr'}$ as it corresponds to many conflicts between the routes.
A detailed derivation of the Hamiltonian mapping can be found in Refs.~\cite{Vikstal_2020_applying, Svensson_2023_Hybrid}.
A vital step in the mapping is reformulating the binary variables $x_r$ into $\pm1$-valued spin variables $s_r$ which are consequently promoted to $Z$ operators.
The derivation in Refs.~\cite{Vikstal_2020_applying, Svensson_2023_Hybrid} define $s_r = 2x_r-1$, issuing a ground state mapping to the original variables as $\ket{0}_r \rightarrow x_r = 1$ and vice versa.
\section{Method}
\label{sec:methods}

Having established the QITE framework and the diagonal Hamiltonian structure of the exact cover and set partitioning problems, we now describe the ansätze and circuit constructions employed in this work.
We begin by reviewing existing QITE ansätze for diagonal Hamiltonians, which serve as baselines for a reduced-parameter construction introduced in Sec.~\ref{sec:reduced_ansatz}.

\subsection{QITE ansätze for diagonal Hamiltonians}
\label{sec:qite_ansatze}

As discussed in Sec.~\ref{sec:prelim_QITE}, the generator of each imaginary-time step is approximated by a parameterized Hermitian operator $\Aop{m}$.
The choice of the operator  basis $\{\sigma_I\}$ in Eq.~\eqref{eq:a_operator} defines the QITE ansatz and determines both the expressivity of the update and the associated quantum and classical resource requirements.

A non-local approximation was introduced in Ref.~\cite{Nishi_Kosugi_Matsushita_2021} by constructing the generator $\Aop{m}$ from combinations of at most $D$-local operators of the $N$-qubit Pauli group. 
Denoting by $\mathcal{P}_N^{(D)}$ the subset
\begin{equation}
\begin{split}
    \mathcal{P}_N^{(D)} =
    \Big\{ \sigma_I = \sigma_1 \sigma_2 \dots \sigma_N \;\big|\;
    \sigma_i \in \{I,X,Y,Z\}, \\
    \sum_i \mathbbm{1}_{\sigma_i \neq I} \leq D
    \Big\},
\end{split}
\end{equation}
the corresponding QITE ansatz expands $\Aop{m}$ over $\mathcal{P}_N^{(D)}$.
Setting $D=N$ recovers the full Pauli basis and the associated exponential scaling, while smaller values of $D$ provide a controlled reduction in the number of parameters.

As discussed in Sec.~\ref{sec:exact_cover}, when starting from a state with real amplitudes in the computational basis, the QITE linear system simplifies and only Pauli strings containing an odd number of $Y$ operators contribute to $\Aop{m}$~\cite{motta_determining_2020}.
Following the notation of Ref.~\cite{xie2025adaptiveweightedqitevqealgorithm}, we denote by $\mathcal{P}_N^{*(D)}$ the restriction of $\mathcal{P}_N^{(D)}$ to Pauli strings with an odd number of $Y$ operators, and refer to the corresponding ansätze as P$D$A. 
That is, P$D$A is the approximation of $\Aop{m}$ constructed with operators from the subset $\mathcal{P}_N^{*(D)}$.

A particularly simple instance is the linear ansatz, P1A, which retains only single-qubit $Y$ operators,
\begin{equation}
    \Aop{m} = \sum_{i=1}^N \acoeff{m}{i} \, Y_i .
\end{equation}
All generators in P1A commute, implying that successive QITE layers can be merged without increasing circuit depth.
In addition, for diagonal Hamiltonians and real-amplitude states, the overlap matrix $\mathbf{S}$ in Eq.~\eqref{eq:qite_setup} reduces to the identity, and the parameter updates depend solely on the vector $\mathbf{b}$.
As a result, P1A admits a purely classical implementation and is equivalent to gradient-descent-based variational approaches~\cite{xie2025adaptiveweightedqitevqealgorithm}.
Despite its simplicity, P1A has been shown to yield competitive results for certain optimization problems, such as weighted and unweighted MaxCut, while failing for others ~\cite{alam_solving_2023, bauer2023combinatorialoptimizationquantumimaginary}.

To introduce entanglement, higher-order ansätze must be considered.
The P2A approximation of $\Aop{m}$ extends P1A by including two-qubit generators of the form $Z_i Y_j$ and $X_i Y_j$ for $i \neq j$, yielding
\begin{equation}
    \Aop{m} = \sum_i \acoeff{m}{i} Y_i
           + \sum_{j \neq i}  \left(\bcoeff{m}{ij} Z_i Y_j + \ccoeff{m}{ij}  X_i Y_j \right) .
\end{equation}
P2A constitutes the lowest-order entangling extension of the linear ansatz and is capable of generating highly entangled states.
However, the number of parameters to be estimated at each QITE step grows like $O(N^2)$, rendering direct implementations prohibitively costly for all but small systems.

\subsection{Reduced-parameter ansatz for dense exact cover instances}
\label{sec:reduced_ansatz}

Here, we introduce a reduced-parameter ansatz tailored to dense exact cover instances, those with a large proportion of non-zero $J_{rr'}$ in Eq.~\eqref{eq:exact_cover_Hamiltonian}, and describe how its structure enables a constant-depth implementation using dynamic quantum circuits.

The approximation given by P2A constitutes the lowest-order entangling extension of P1A; however, the number of parameters $\acoeff{m}{I}$ to be estimated at each QITE step $m$ scales rapidly with system size.
With a Suzuki-Trotter decomposition of $\Aop{m}$, a general QITE circuit may be written as a layered product of parameterized unitaries,
\begin{equation}
    \label{eq:qite_layered_product}
    U_{\text{QITE}}
    =
    \prod_{m=1}^{M} \prod_I
    e^{- i \acoeff{m}{I} \sigma_I}.
\end{equation}
The realizable unitaries $U_{\text{QITE}}$ are given by its dynamical Lie group, which in turn is determined by the dynamical Lie algebra (DLA) of the gate generators $\mathcal{S}=\{\sigma_I\}$~\cite{DAlessandro_2021}.
The vector space spanned by the nested commutators of $i\mathcal{S}$ defines the DLA.
In the construction of P2A, we consider gate generators drawn from $\mathcal{P}_N^{*(2)}$. 
Spanning all possible nested commutators of $i\mathcal{P}_N^{*(2)}$ for $N\geq 4$ leads to the DLA $\mathfrak{so}(2^N)$~\cite{Wiersema_Kokcu_Kemper_Bakalov_2024}.
Indeed, the Pauli operators with an odd number of $Y$ operators are purely imaginary and lead to real antisymmetric generators after multiplying by $i$.

For a general QITE circuit, the idea is to approximate the normalized evolution of ITE as closely as possible, which requires the operator basis of $\Aop{m}$ to be large enough to contain the correlations of the evolution \cite{motta_determining_2020}.
However, for ground state preparation it is not necessary to approximate ITE faithfully at every intermediate step.
It suffices that the layered circuit can eventually reach the same low-energy manifold as the exact ITE.
This viewpoint aligns with VarQITE approaches, which deliberately restrict the reachable Hilbert space by using a parametrized quantum circuit ansatz~\cite{McArdle_2019, Kolotouros_2025, Benedetti_2021_hardware, wang2023symmetryenhancedvariationalquantum, Gacon_Variational_2024}.
Consequently, the operator basis of $\Aop{m}$ need not span $\mathfrak{so}(2^N)$; it must only generate a sufficiently expressive ansatz.

\paragraph{Removal of $X_iY_j$ operators.}

We first restrict P2A by removing all $X_iY_j$ generators.
Note that the resulting operator set $\{ Y_i,\, Z_i Y_j \mid i,j=1,\dots,N \}$ still produces a DLA with $X_iY_j$ operators from the first commutator level of the DLA construction.
Thus, eliminating the gate generators $X_iY_j$ does not restrict the kind of accessible  parametrized unitaries, but may increase the number of layers required.
The trade-off is therefore between per-layer parameter count and circuit depth.

\paragraph{Restriction to a single pivot qubit.}

We further reduce the operator set by selecting a single pivot (control) qubit $k$ and retaining only operators of the form
\begin{equation}
    \label{eq:qite_reduced_generators}
    \mathcal{S}_{\mathrm{red}}
    =
    \{\, Y_i, Y_k,\; Z_k Y_i \mid i=1,\dots,N,\; i\neq k \,\}.
\end{equation}
This reduces the number of two-qubit operators from $O(N^2)$ to $O(N)$ while preserving the ability to entangle qubit $k$ with the remainder of the system. That is, we have a connected coupling graph, with all the qubits linked via the pivot qubit.

We note that the DLA generated by $i\mathcal{S}_{\mathrm{red}}$ is strictly contained in $\mathfrak{so}(2^N)$, since its dimension is strictly smaller than $\text{dim}{(\mathfrak{so}(2^N))}$.
In particular, it cannot generate operators of the form $X_\ell Y_i$ for $\ell \neq k$, which is required as $\mathfrak{so}(2^N)$ contains all Pauli strings with an odd number of $Y$ operators.
With only $Y$ as valid single qubit rotation, a universal real amplitude set requires controlled operations on each qubit.
One such set for $N\geq4$ is 
\begin{equation}
    \label{eq:diagonal_qite_universal_generators}
    \{\, Y_i,\; Z_k Y_{k+1 \bmod N}
      \mid i,k = 1,\dots,N \,\}.
\end{equation}
Consequently, the expressivity of the circuit is reduced for $\mathcal{S}_{\mathrm{red}}$.
Nonetheless, limiting expressivity may be advantageous, given its close connection to the emergence of barren plateaus~\cite{Holmes_2022_connecting, Larocca2022diagnosingbarren, Ragone_2024}.
Even though QITE is not an exact gradient descent setup, as $b_I \propto \nabla_{a_I} \bra{\psi}H\ket{\psi} $ goes to 0 $\forall I$, the linear system will either be highly underdetermined or have an all zero solution.

\paragraph{Dense-instance motivation.}

For dense exact cover instances, the preference of $\mathcal{S}_{\mathrm{red}}$ compared to Eq.~\eqref{eq:diagonal_qite_universal_generators} is motivated by the structure of the low-energy manifold.
Starting from the equal superposition state $\ket{+}^{\otimes N}$, ITE converges to an equal superposition of the lowest-energy eigenstates.
Qubits which take identical values across all these states are not entangled with the rest of the system.
As an illustrative example, for
\begin{align}
    \label{eq:ground_state_structure}
    \ket{\psi}
    &=
    \tfrac{1}{\sqrt{2}}\bigl(\ket{001111}+\ket{110011}\bigr) \nonumber \\
    &=
    \tfrac{1}{\sqrt{2}}\bigl(\ket{0011}+\ket{1100}\bigr)\otimes\ket{11},
\end{align}
the last two qubits are disentangled from the first four.
In such situations, removing operators that couple these qubits to the remainder does not restrict access to the ground state.
Dense instances empirically exhibit a large proportion of variables with this property, suggesting that a single appropriately chosen pivot qubit $k$, which is $\ket{0}$ in at least one of the ground states, can suffice to find one of the lowest energy computational basis states.
The generators of Eq.~\eqref{eq:diagonal_qite_universal_generators} on the other hand, will entangle an unnecessarily large proportion of qubits for ground states with structure as in Eq.~\eqref{eq:ground_state_structure}.
On that account, our tests show they are not preferred even though the generators allow a constant unitary depth circuit, assuming any topology where a cycle can be created for the $Z_NY_1$ operation.

\paragraph{Choice of pivot qubit.}
Selecting an optimal pivot qubit $k$ requires knowledge of the solution.
In this work, $k$ is chosen from a variable participating in a known ground-state configuration obtained via a classical solver, in order to demonstrate the effectiveness of the reduced ansatz when a suitable choice is made.
To remove this requirement of prior-knowledge, we introduce a simple heuristic for selecting $k$, detailed in Appendix~\ref{sec:app_heur}.
Numerical tests indicate a small number of expected restarts.
0.39 on average, and 7 in the worst case for problem sizes up to $120$ qubits.

\paragraph{Final reduced ansatz.}

With these restrictions, the QITE operator at layer $m$ is
\begin{equation}
    \label{eq:final_reduced_ansatz}
    \Aop{m}
    =
    \sum_{i=1}^{N} \acoeff{m}{i}\, Y_i
    +
    \sum_{\substack{i=1 \\ i\neq k}}^{N}
    \bcoeff{m}{i}\, Z_k Y_i.
\end{equation}
Compared to P2A, this reduces the number of parameters and quantum gates per layer from $O(N^2)$ to $O(N)$.

\subsection{Layer compression via step merging}
\label{sec:step_merging}
In the following section, we describe how this reduced ansatz structure enables layer compression and a constant-depth circuit implementation using dynamic quantum circuits.
The QITE circuit after $M$ iterations consists of a product of layered unitaries,
\begin{equation}
    U_{\mathrm{QITE}}
    =
    \prod_{m=1}^{M}
    e^{-i \Delta\tau \Aop{m}},
\end{equation}
where $\Aop{m}$ denotes the operator defined in Eq.~\eqref{eq:final_reduced_ansatz}.
Even if each $\Aop{m}$ is sparse in the chosen operator basis, the sequential application of $M$ layers increases the circuit depth linearly in $M$.

To mitigate this growth, we employ the step-merging (or compression) procedure introduced in Ref.~\cite{Nishi_Kosugi_Matsushita_2021}.
Using a first-order reverse Suzuki–Trotter expansion, the product of layers may be approximated as
\begin{equation}
    \label{eq:step_merging}
    \prod_{m=1}^{M}
    e^{-i \Delta\tau \Aop{m}}
    =
    \exp\!\left(
        -i \Delta\tau
        \sum_{m=1}^{M} \Aop{m}
    \right)
    +
    \mathcal{O}(\Delta\tau^2),
\end{equation}
where the error arises from non-commuting terms in different layers.

Specifically, the approximation error is governed by the commutators $[\Aop{m},\Aop{m'}]$.
If the operator set contains many non-commuting operators, these commutators accumulate and the compression error increases.
Conversely, reducing the operator set lowers the number of non-vanishing commutators and therefore reduces the error introduced by step merging.

For the reduced ansatz~\eqref{eq:final_reduced_ansatz}, the number of distinct non-commuting operator pairs is significantly smaller than in the full P2A construction.
As a result, the compressed operator
\begin{equation}
    A_{\mathrm{tot}}
    =
    \sum_{m=1}^{M} \Aop{m}
\end{equation}
provides a controlled first-order approximation to the layered circuit while keeping the quantum depth independent of $M$.

Importantly, step merging does not alter the parameter-update rule of QITE.
It only modifies the physical realization of the accumulated unitary on hardware.
The classical update of parameters remains sequential in $M$, while the quantum circuit implements the compressed operator corresponding to the current cumulative parameters.

\subsection{Constant-entangling-depth implementation via dynamic fan-out circuits}
\label{sec:dynamic_fanout}

The reduced ansatz introduced in Sec.~\ref{sec:reduced_ansatz} and compressed via step merging in Sec.~\ref{sec:step_merging} yields a single effective unitary of the form
\begin{equation}
    U_{\mathrm{comp}}
    =
    \exp\!\left(
        - i \Delta\tau
        \sum_{i=1}^{N} \alpha_i Y_i
        - i \Delta\tau
        \sum_{\substack{i=1 \\ i\neq k}}^{N} \beta_i Z_k Y_i
    \right),
\end{equation}
where the coefficients $\alpha_i$ and $\beta_i$ are accumulated from the QITE iterations.
This operator consists exclusively of single-qubit rotations and two-qubit interactions coupling a single control qubit $k$ to all other qubits.

\paragraph{Entangling-depth considerations.}

\begin{figure}[hbtp]
    \centering

    \makebox[0.45\textwidth][l]{(a)}%
    \par\vspace{-1.2em}
    \begin{minipage}{0.45\textwidth}
        \centering
        $$\Qcircuit @C=1em @R=0.5em { 
        & \gate{R_Y(\theta_1)} & \multigate{3}{\displaystyle{\prod^n_{\substack{i=0,\\i\neq k}}}R_{Z_kY_i}(\theta_{k,i})} & \qw \\ 
        & \gate{R_Y(\theta_2)} & \ghost{\displaystyle{\prod^n_{\substack{i=0,\\i\neq k}}}R_{Z_kY_i}(\theta_{k,i})} & \qw \\
        & \cdots               & \nghost{U}                                                    &  \\
        & \gate{R_Y(\theta_n)} & \ghost{\displaystyle{\prod^n_{\substack{i=0,\\i\neq k}}}R_{Z_kY_i}(\theta_{k,i})} & \qw 
        }$$
    \end{minipage}
    \hspace{1em}
    \makebox[0.45\textwidth][l]{(b)}%
    \par\vspace{-1.2em}
    \begin{minipage}{0.45\textwidth}
        \centering
        $$\Qcircuit @C=1em @R=.4em { 
        & \qw & \multigate{4}{\displaystyle{\prod^n_{\substack{i=0,\\i\neq k}}}R_{Z_kY_i}(\theta_{k,i})} & \qw & & & & \ctrl{4} & \qw & \ctrl{4} & \qw\\           
        & \qw & \ghost{\displaystyle{\prod^n_{\substack{i=0,\\i\neq k}}}R_{Z_kY_i}(\theta_{k,i})} & \qw & & & & \targ & \gate{R_Y(\theta_{k,2})} & \targ & \qw\\
        &     &                                             &     & \push{\rule{0.1em}{0em}=\rule{0.1em}{0em}} & \\
        & \cdots & \nghost{U} &  & & & &  & \cdots &  & \\
        &\qw & \ghost{\displaystyle{\prod^n_{\substack{i=0,\\i\neq k}}}R_{Z_kY_i}(\theta_{k,i})} & \qw & & & & \targ & \gate{R_Y(\theta_{k,n})} & \targ & \qw
        }$$
    \end{minipage}

    \caption{
        (a) Single layer of the QITE ansatz.
        (b) Implementation of a constant‑depth parameterized multi‑qubit rotation using two fan-out gates \cite{baumer_measurement-based_2024} and $k=1$.
    }
    \label{fig:quantum_circuits}
\end{figure}
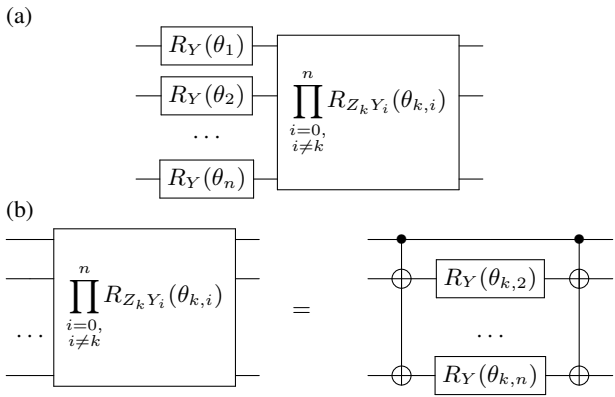

In a purely unitary circuit model with restricted connectivity, implementing the $Z_k Y_i$ rotations for all $i$ requires mediating interactions between qubit $k$ and the remaining $N-1$ qubits.
On a one-dimensional or nearest-neighbor architecture, this necessitates a SWAP network, resulting in an entangling-gate depth that scales linearly with system size.
Even in more connected topologies, the circuit depth associated with the two-qubit gates scales as $O(N)$, as interactions must be applied sequentially \cite{ogorman2019generalizedswapnetworksnearterm}.
Dynamic quantum circuits allow for this limitation to be circumvented.
By combining mid-circuit measurements with classical feed-forward, multi-qubit fan-out operations, which distributes the logical state of a control qubit onto multiple target qubits, can be realized with constant two-qubit entangling gate depth~\cite{baumer_measurement-based_2024}.
Once the fan-out is established, all rotations of the form $\prod^n_{\substack{i=0,\\i\neq k}}R_{Z_k Y_i}(2\beta_i)$ can be applied in parallel, after which the fan-out is reversed.
As a result, the entire block of two-qubit interactions is implemented using a constant number of entangling layers, independent of $N$.
Figure~\ref{fig:quantum_circuits}(b) illustrates this construction for a single layer, which is the complete circuit when compressing.

\begin{figure}[hbtp]
    \centering
    $$\Qcircuit @C=1em @R=.4em{
    &\gate{R_Y(\theta_1)} & \meter &\cw \cwx[1] & \cw & \cw \cwx[1] & \cw \\
    & \qw& \gate{R_Y(\theta_2)} & \gate{X} & \gate{R_Y(\theta_{1,2})} & \gate{X} & \qw\\
    & \cdots& \cdots & \cwx[1] \cwx[-1] & \cdots & \cwx[1] \cwx[-1] & \cdots \\
    & \qw& \gate{R_Y(\theta_n)} &\gate{X} &  \gate{R_Y(\theta_{1,n})} & \gate{X} & \qw}$$
    \caption{Circuit layer of the semi-classical approach with $k=1$.
    This simplification of Fig. \ref{fig:quantum_circuits} is possible when qubit $k$ only has diagonal operations after the last fan-out gate.
    The double lines denote the flow of classical information.}
    \label{fig:semi-classical_circuit}
\end{figure}
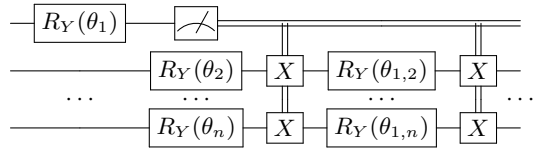

\paragraph{Semi-classical implementation.}
A further reduction in both quantum and classical resources is possible in cases where the control qubit $k$ is only subject to diagonal operations after the QITE layer.
In this situation, the $Z$ measurement on qubit $k$ can be commuted to immediately after its single-qubit rotation, yielding a semi-classical fan-out circuit.
This implementation requires no entangling gates and only a single mid-circuit measurement, at the expense of restricting the allowed measurement bases in subsequent QITE steps, depicted in Fig. \ref{fig:semi-classical_circuit}.
As QITE generally requires measurements in all Pauli bases, this approach is not universally applicable, but can be advantageous in specific regimes.

\paragraph{Scope of the depth reduction.}

It is important to emphasize that dynamic circuits do not reduce the logical complexity of the operation being implemented.
Rather, they exchange quantum coherence time for measurement and classical processing.
In the absence of mid-circuit measurement and feed-forward, the same interaction pattern can only be realized by sequentially routing the control qubit through the system, leading to linear entangling-gate depth.
Thus, the depth advantage obtained here is intrinsically tied to the availability of dynamic circuit capabilities.

We stress that the constant-depth implementation relies critically on the restricted operator structure of the reduced ansatz.
For more general QITE ansätze containing interactions between arbitrary qubit pairs, dynamic fan-out alone does not suffice to achieve constant entangling-gate depth.
Namely, a small number of pivot qubits are required.
In this sense, the depth reduction is not generic to QITE, but is enabled by the combination of (i) parameter reduction and (ii) dynamic circuit primitives.

Table \ref{tab:resource_scale} presents resource requirements of a single layer of the circuits when applied to a problem of size $N$.
While the total number of two-qubit gates increase for the dynamic implementation compared to the unitary circuit, the reduction in entangling-gate depth directly targets the dominant decoherence channel on current hardware.
The total runtime of the dynamic circuit is shorter than the unitary whenever the time for two mid-circuit measurements and feedback operations and the $10$ CNOT gates are shorter than the time of $3N-4$ CNOT gates.
With reported times from \verb|ibm_pittsburgh|, $t_\text{measure}=1.288\times 10^{-6}, t_{\text{feedback}}=0.6 \times 10^{-6}$ and $t_{\text{CNOT}} = 88\times 10^{-9}$, the dynamic circuit has a shorter runtime for $N>18$.
For the semi-classical implementation which requires only one mid-circuit measurement, this happens at $N>15$ instead.

\begin{table}[h]
    \centering
    \caption{\label{tab:resource_scale}Resource scaling of unitary, dynamic and semi-classical implementation.
    The unitary approach uses a SWAP-network rather than two unitary fan-out gates, where a SWAP-gate is assumed to be constructed by 3 CNOT gates.
    Allowing more complex hardware topologies, the minimum CNOT depth for the unitary is still $2N$, as each rotation gate requires two CNOT gates each.}
    \begin{ruledtabular}
    \begin{tabular}{l c c c}
        \textbf{Implementation} & \textbf{Unitary} & \textbf{Dynamic} & \textbf{Semi-classical}  \\ \midrule
        Connectivity & 1D-line & 1D-line & Any \\
        \# Qubits & $N$ & $2N-1$ & $N$  \\
        CNOT Depth & $3N-4$ & 10 & 0  \\
        \# CNOTs & $3N-4$ & $6N-8$ & 0 \\
        \# Mid-circuit meas. & 0 & $2N-2$ & 1\\
        \# Classical comp. & 0 & $\lfloor \frac{3}{4}N\rfloor$ & 0\\
    \end{tabular}
    \end{ruledtabular}
\end{table}

\subsection{Classical simulation techniques}
Even though current quantum hardware is noisy, their inherent capability of simulating quantum systems is unmatched.
Modern computers are capable of matrix multiplications of the size $10^7 \times 10^7$ within reasonable time, allowing exact state-vector simulations up to 26 qubits.
With specialized hardware, the number is pushed to $\sim40$ qubits \cite{Lykov_2023_fast} and possibly further with approximate tensor-methods, dependent on the entanglement structure \cite{Shi_2006_classical, pang2020efficient2dtensornetwork}.
Other problem-specific techniques for simplifying simulations include stabilizers \cite{Aaronson_improved_2004}, group theoretic \cite{goh2025liealgebraicclassicalsimulationsquantum}, constant-depth specific \cite{Bravyi_2021} and low-weight Pauli propagation \cite{angrisani2025classicallyestimatingobservablesnoiseless, Shao_2024_simulating},
but the P$D$A-ansätze of QITE with $D>1$ does not fall into any of these categories. 
The ansatz has the possibility of creating moderately entangled states, ruling out tensor-networks.
It also requires measurements of observables consisting of all types of Pauli operators, giving an exponential scaling of the $\mathfrak{g}$-sim method \cite{goh2025liealgebraicclassicalsimulationsquantum}.
Additionally, the circuit depth (information light cone) increases polynomially with $D$ and $N$ ruling out constant-depth methods.
Finally, the circuit only has parameterized gates around the $Y$-axis, and the starting parameters are not sampled but rather set to zero, which are requirements for low-weight Pauli propagation.
The approach of Ref. \cite{Lykov_2023_fast} could be applied, as only diagonal problems are considered, but it is still exponential in scaling and heavily reliant on computational power.
Generally, noise allows for easier classical approximations; however, there is a threshold where these simulation methods become too expensive, and the remaining possibility is testing with real quantum hardware.

\section{Results}
\label{sec:results}
This section presents results from noiseless simulations on exact cover and set partitioning instances, hardware experiments comparing the three aforementioned circuit implementations, and an analysis of the hardware requirements relevant for dynamic circuit constructions.
The problem instances come from two separate sets, the first studied in ~\cite{Vikstal_2020_applying, Willsch_2022} and the second in ~\cite{Svensson_2023_Hybrid, cacao_2024_the}, available at ~\cite{marika_instances}. 
All instances have been extracted from real-world tail assignment problems with a simplified cost structure. 

\subsection{Exact cover}
We begin by applying the linear QITE ansatz to the exact cover instances studied in order to identify instances for which the linear ansatz fails to converge to a valid solution.
These instances serve as benchmarks for assessing the impact of the reduced-parameter ansatz from Sec~\ref{sec:reduced_ansatz}.

The metric of solution quality is called success probability, $p$(gs), defined as the probability of measuring any of the minimum eigenvalue basis states of the problem Hamiltonian 
\begin{equation}
    \psuccess = \sum_{x\in GS} |\langle x| \psi\rangle|^2.
\end{equation}
$GS$ is the set of all computational basis states yielding the lowest eigenvalue of the Hamiltonian, and $\ket{\psi}$ is the final state produced by the algorithm.
The success probability is chosen as metric of solution quality due to exact cover being a feasibility/constraint satisfaction problem --- only ground states are valid solutions, it is not sufficient to find low-energy states.
An instance is defined as unsolved whenever the success probability is less than $1\%$.
In the context of fault-tolerant quantum algorithms, a constant $1\%$ overlap with the target eigenstate is sufficient for polynomial-resource eigenstate preparation and amplification~\cite{Lee_2023, Aspuru_Guzik_2005, zhang2025quantumimaginarytimeevolutionpolynomial}.
Moreover, sampling from a state with constant overlap can itself be performed in polynomial time~\cite{abbas_challenges_2024}.

The results obtained with the linear QITE ansatz are reported in Appendix~\ref{app:linear_QITE_performance}.
Generally, the performance is tightly connected to the density of the instance, with a sparse constraint structure often resulting in a correct solution.

Applying the proposed reduced-parameter, higher-order ansatz to the most difficult instances, we observe a substantial improvement in solution quality.
To ensure the simplifications proposed in Sec.~\ref{sec:reduced_ansatz} are just, we further compare against variants employing $XY$ rotations, no layer compression, and the full set of $ZY$ rotations.

\begin{figure}
    \centering
    \includegraphics[width=1\linewidth]{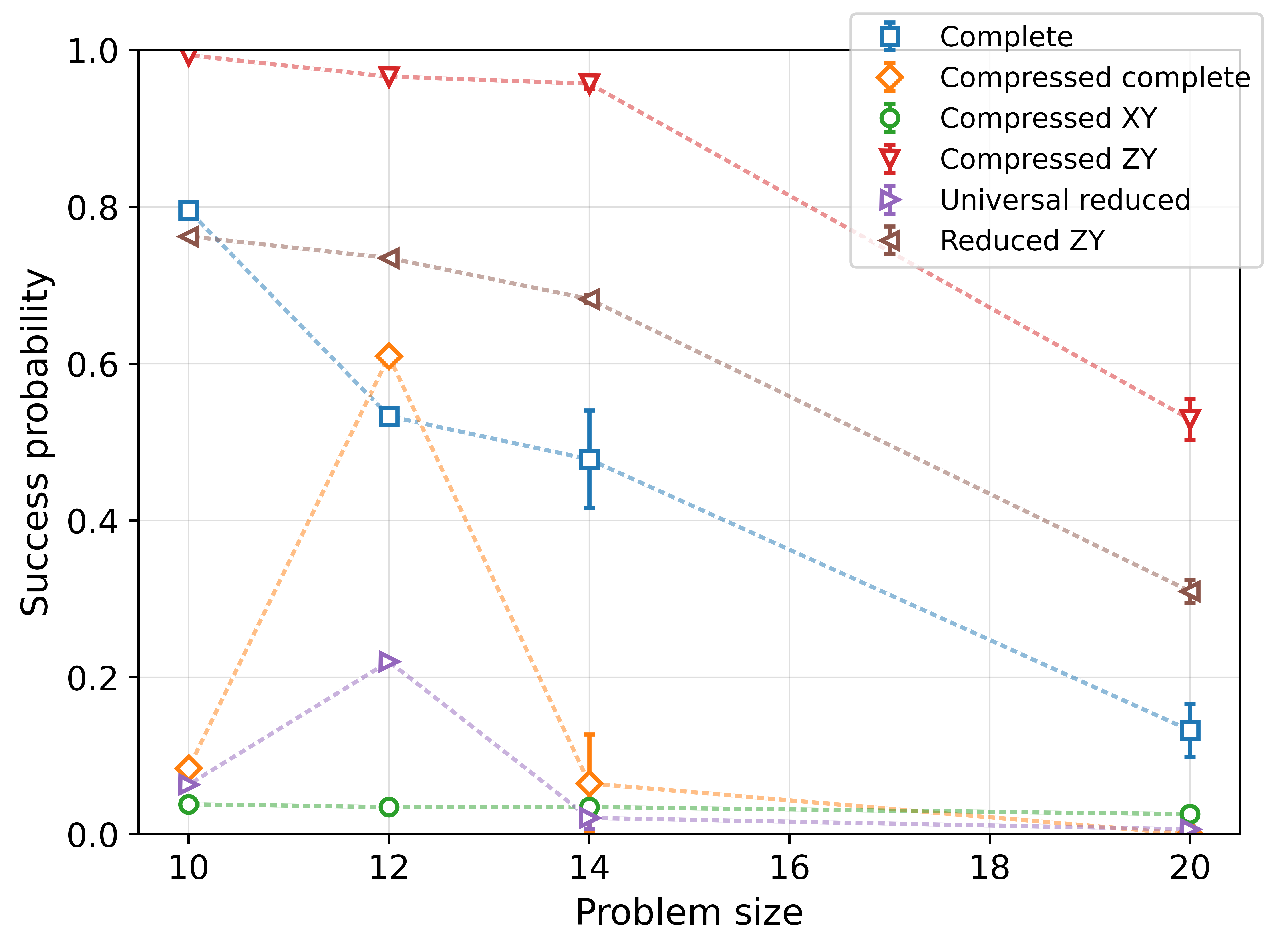}
    \caption{Mean success probabilities for difficult exact cover instances ($M=125$, $\Delta\tau=0.004$) across different variants of P2A. 
    Error bars indicate the variance, $\sigma^2$ over instances. 
    ``Reduced ZY'' refers to the set $\mathcal{S}_{\mathrm{red}}$, and ``Compressed ZY'' is when compressing the circuit.
    ``Compressed XY'' is when utilizing $XY$ rotations instead of $ZY$.
    ``Complete'' is the original ansatz P2A.
    ``Universal reduced'' is the set of Eq.~\eqref{eq:diagonal_qite_universal_generators}.}
\label{fig:exact_cover_p2a_ansatz}
\end{figure}
As shown in Fig.~\ref{fig:exact_cover_p2a_ansatz}, the proposed ansatz introduced in this work achieves the highest success probability across all system sizes considered, while simultaneously reducing circuit depth and parameter count.
Notably, layer compression further improves performance in these instances; a discussion of this effect is provided in Appendix~\ref{app:compression}.

\subsection{Set partitioning}
To demonstrate the proposed approach extends beyond pure feasibility problems, we reintroduce the objective function in Eq.~\eqref{eq:set_partitioning} and solve the full set partitioning problem. 
The penalty parameters are chosen following Ref.~\cite{cacao_2024_the}, which guarantees the rescaled Hamiltonian preserves the ground-state solution of the original problem formulation, while yielding a tighter bound than the sum of all objective value coefficients.

Introducing penalty terms to the Quadratic Unconstrained Binary Optimization (QUBO) formulated Hamiltonian increases the magnitudes of the coefficients, requiring smaller imaginary-time steps.
Moreover, the energy landscape of the Hamiltonian becomes increasingly complicated and contains additional local minima~\cite{Svensson_2023_Hybrid}, reflecting the general difficulty of constrained optimization problems encoded as QUBOs for quantum algorithms.

\begin{figure}
    \centering
    \includegraphics[width=1\linewidth]{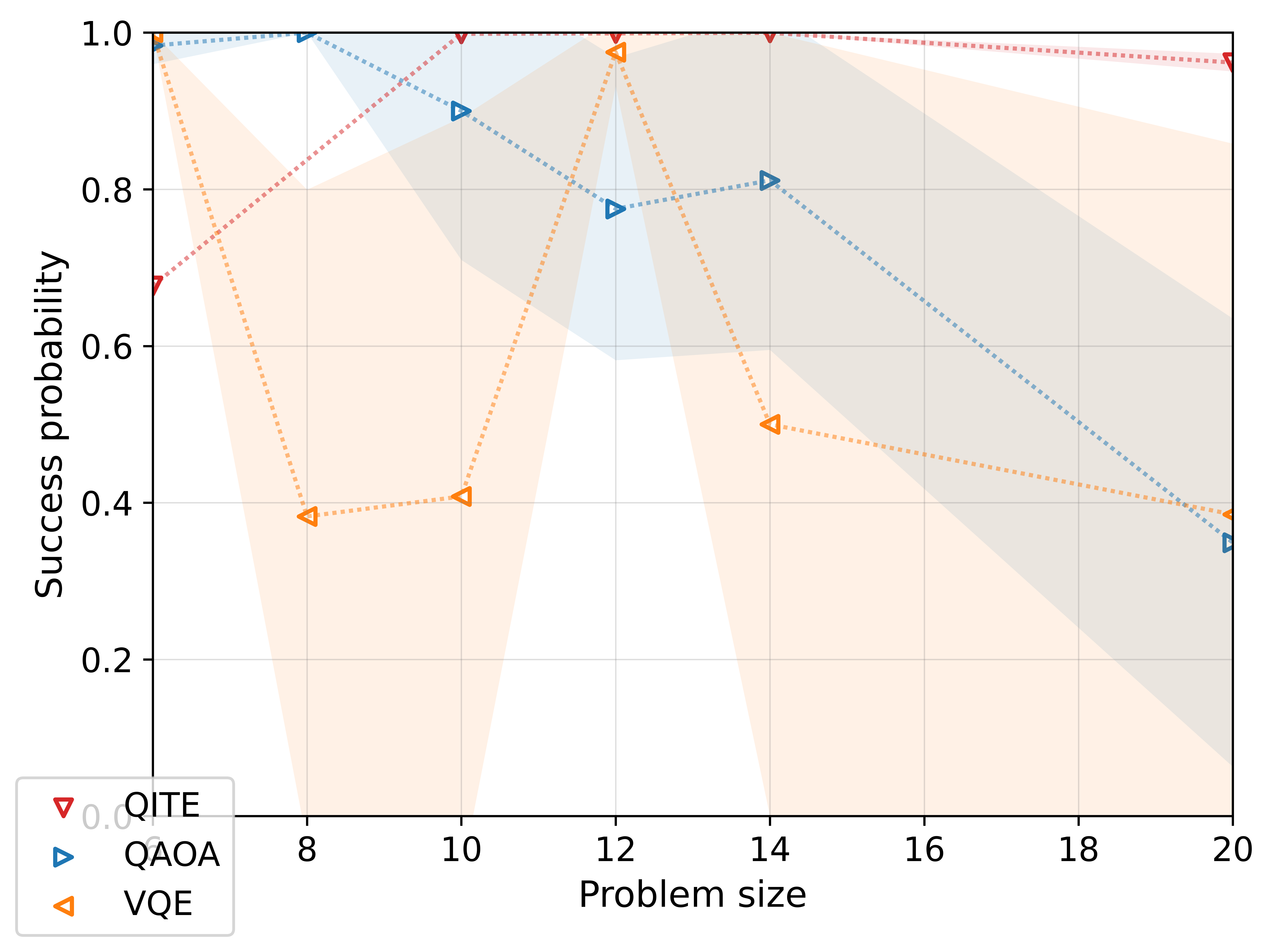}
    \caption{Mean success probabilities of the reduced parameter QITE on the difficult set partitioning instances using $\Delta\tau = 8\times10^{-5}, M = 100$.
    The results are compared to QAOA and VQE results from \cite{Svensson_2023_Hybrid,cacao_2024_the}
    The shaded area gives the variance, $\sigma^2$, which is notably low for QITE.}
    \label{fig:pgs_of_size_setPart}
\end{figure}

As in the exact cover case, we first apply P$1$A to identify difficult instances (see Appendix~\ref{app:linear_QITE_performance}). 
Instances that remain unsolved, $\psuccess<0.01$, are then revisited using the reduced-parameter ansatz.
Using the same problem instances as Refs.~\cite{cacao_2024_the, Svensson_2023_Hybrid}, we compare the performance of the reduced-parameter QITE ansatz to previously reported QAOA and VQE results.
Figure~\ref{fig:pgs_of_size_setPart} shows the average success probabilities obtained for difficult set partitioning instances.
In addition to the average performance, the variance across instances is reported. 
Notably, the variance $\sigma^2$ observed for the reduced parameter QITE is small across all problem sizes considered.
The depth of the QAOA circuits is $p=20$ for the 20-variable instances and $p=50$ otherwise.
Using $100$ QITE iterations ($M$) is justified by the constant circuit depth and the polynomial scaling of the classical resources between iterations.

\subsection{Hardware experiments}
To test the dynamic circuit capabilities in current quantum hardware, we tested the three different implementations described in Sec.~\ref{sec:methods} on exact cover instances of sizes 6, 10, 12, 14, 20, and 30 qubits using IBM superconducting quantum devices, with the compressed reduced-parameter QITE ansatz.

For problem sizes between 6 and 14 qubits, each instance was run for as many iterations as required for the noiseless simulation to reach a ground-state success probability of $2/3$, which we set as the acceptance rate.
The imaginary-time step sizes $\Delta \tau$ were chosen adaptively such that the largest parameter value after the first iteration remained below $\pi/4$.
This ensures the single qubit $R_Y$-rotations are kept moderate after the first iteration.
The resulting hardware performance is summarized in Table~\ref{tab:hardware_results_error}, in which we report the average of the absolute energy difference between each implementation and the noiseless benchmark. 
Table~\ref{tab:hardware_results_pgs} shows the success probabilities of the hardware tests, with the semi-classical and unitary approaches reaching the highest values.

Despite small absolute errors in expectation value from the unitary and semi-classical implementations, the final success probability is far from the target $2/3$.
To mitigate hardware errors, specifically shot noise, the 20-qubit instance was therefore executed with the smallest possible step size $\Delta\tau$ that still allowed the success probability to reach at least $1\%$ within 20 iterations.
The 30-qubit instance was executed for 30 iterations with $\Delta\tau = 0.003$, as no corresponding noiseless simulation reference was available.
Further experimental details are provided in Appendix~\ref{sec:exp_details}.

\begin{table}[htbp]
\centering
\caption{
\label{tab:hardware_results_error} 
Mean and variance of the absolute difference $|E_{\text{exp}}-E_{\text{sim}}|$ between the energy expectation values of the different experiments and noiseless simulations across $M$ iterations for one instance of each size.
Energy is normalized as $E=(\langle H \rangle-E_{\text{min}})/(E_{\text{max}}-E_{\mathrm{min}})$. We consider instance sizes $N$, number of iterations $M$, and imaginary-time step size $\Delta\tau$.}
\begin{ruledtabular}
\begin{tabular}{ccc c c c} 
    $N$& $M$ & $\Delta\tau$& \textbf{Dynamic} & \textbf{Unitary} & \textbf{Semi-classical}  \\ \midrule
    6  & 9 &$0.040$ & $0.17 \pm 0.01$ & $0.05\pm 0.02$ & $0.04\pm 0.03$ \\
    10 &20 &$0.021$ & $0.17\pm 0.01$ & $0.04\pm 0.06$ & $0.02\pm 0.04$ \\
    12 & 30 & $0.017$& $0.37\pm 0.01$ & $0.03\pm 0.05$ & $0.03\pm 0.07$ \\
    14 &42 &$0.012$ & $0.39\pm 0.01$ & $0.02\pm 0.02$ & $0.01\pm 0.01$  \\ 
\end{tabular}
\end{ruledtabular}
\end{table}

\begin{table}[htbp]
\centering
\caption{
\label{tab:hardware_results_pgs} 
Mean and variance of the success probability $\psuccess$ achieved by the different implementations at the final step, for instance sizes $N$.
The number of iterations $M$, and imaginary-time step size $\Delta\tau$ are left out as they differ between instances. 
The target success probability for convergence is $2/3$.}
\begin{ruledtabular}
\begin{tabular}{c c c c} 
    $N$& \textbf{Dynamic} & \textbf{Unitary} & \textbf{Semi-classical}  \\ \midrule
    6   & $0.40 \pm 0.019$ & $0.613 \pm 0.012$ & $0.4889 \pm 0.009$ \\
    10 & $0.021 \pm 0.011$ & $0.171 \pm 0.026$ & $0.250 \pm 0.013$ \\
    12 &  $0.009 \pm 0.004$ & $0.215 \pm 0.033$ & $0.114 \pm 0.014$ \\
    14& $0.004 \pm 0.003$ & $0.153 \pm 0.015$ & $0.271 \pm 0.008$  \\ 
\end{tabular}
\end{ruledtabular}
\end{table}

\begin{figure}[htbp]
    \centering
    \includegraphics[width=1\linewidth]{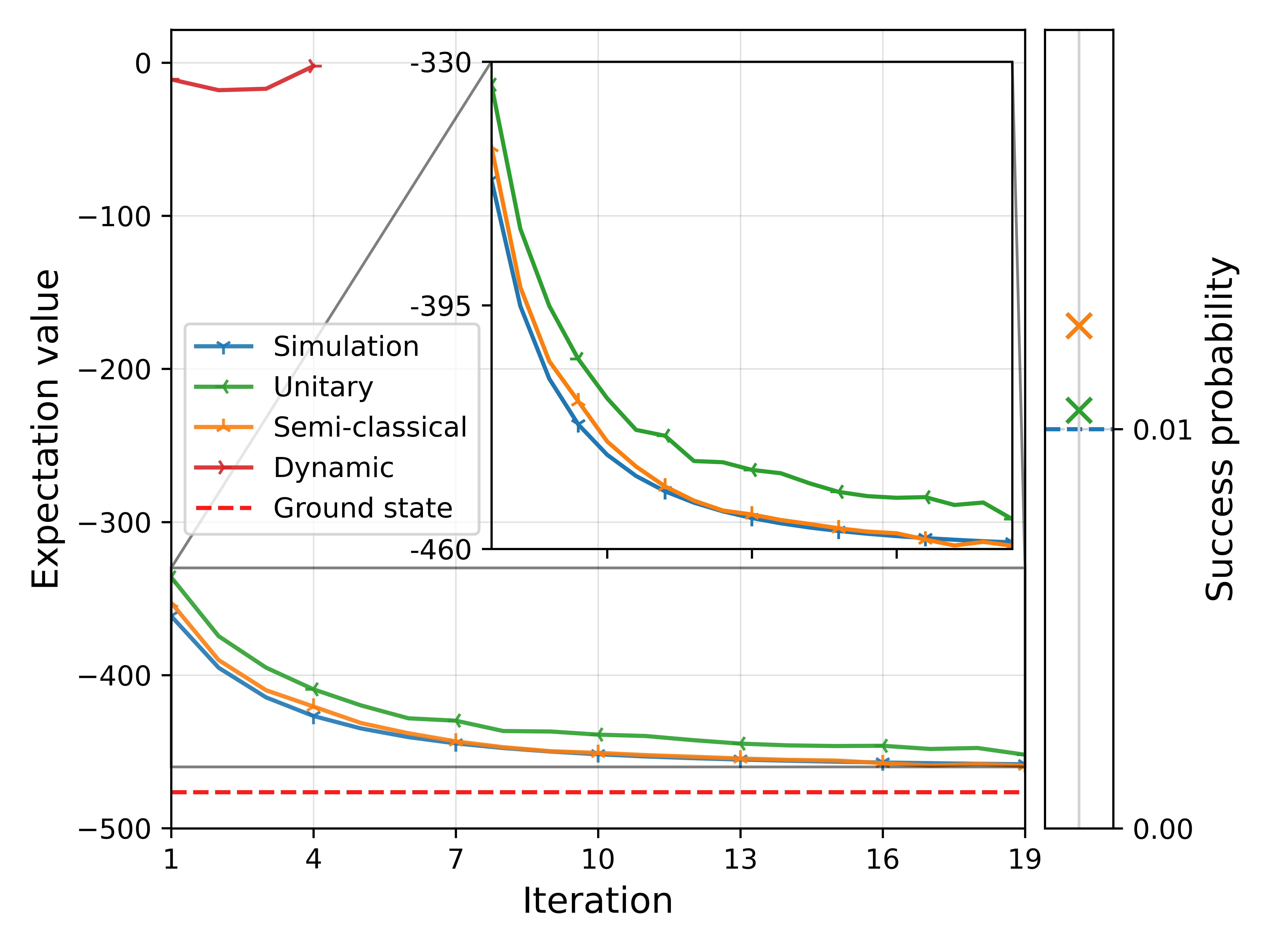}
    \caption{Expectation value of the problem Hamiltonian for a 20-qubit exact cover instance obtained from noiseless simulation and from unitary (with and without semi-classical simplification) and dynamic circuit hardware implementations, using $M=20$ and $\Delta\tau=0.005$.
    The lines start at iteration $m=1$ of QITE, as all methods yield zero expectation value before the first iteration.
    The dashed red line indicates the ground-state energy.
    The crosses indicate samples taken at iteration $m=20$, corresponding to the measured overlap with the ground state.}
    \label{fig:hardware20qb}
\end{figure}

Figure~\ref{fig:hardware20qb} compares the expectation value of the exact cover Hamiltonian for the different implementations across 20 iterations for the 20-qubit instance in \verb|ibm_pittsburgh|.
As for the smaller instances, the dynamic implementation performs worst.
In this case, the dynamic run was terminated after four iterations, as the energy increased beyond the initial value.
The unitary and semi-classical implementations exhibit similar behavior throughout the run.

\begin{figure}
    \centering
    \includegraphics[width=1\linewidth]{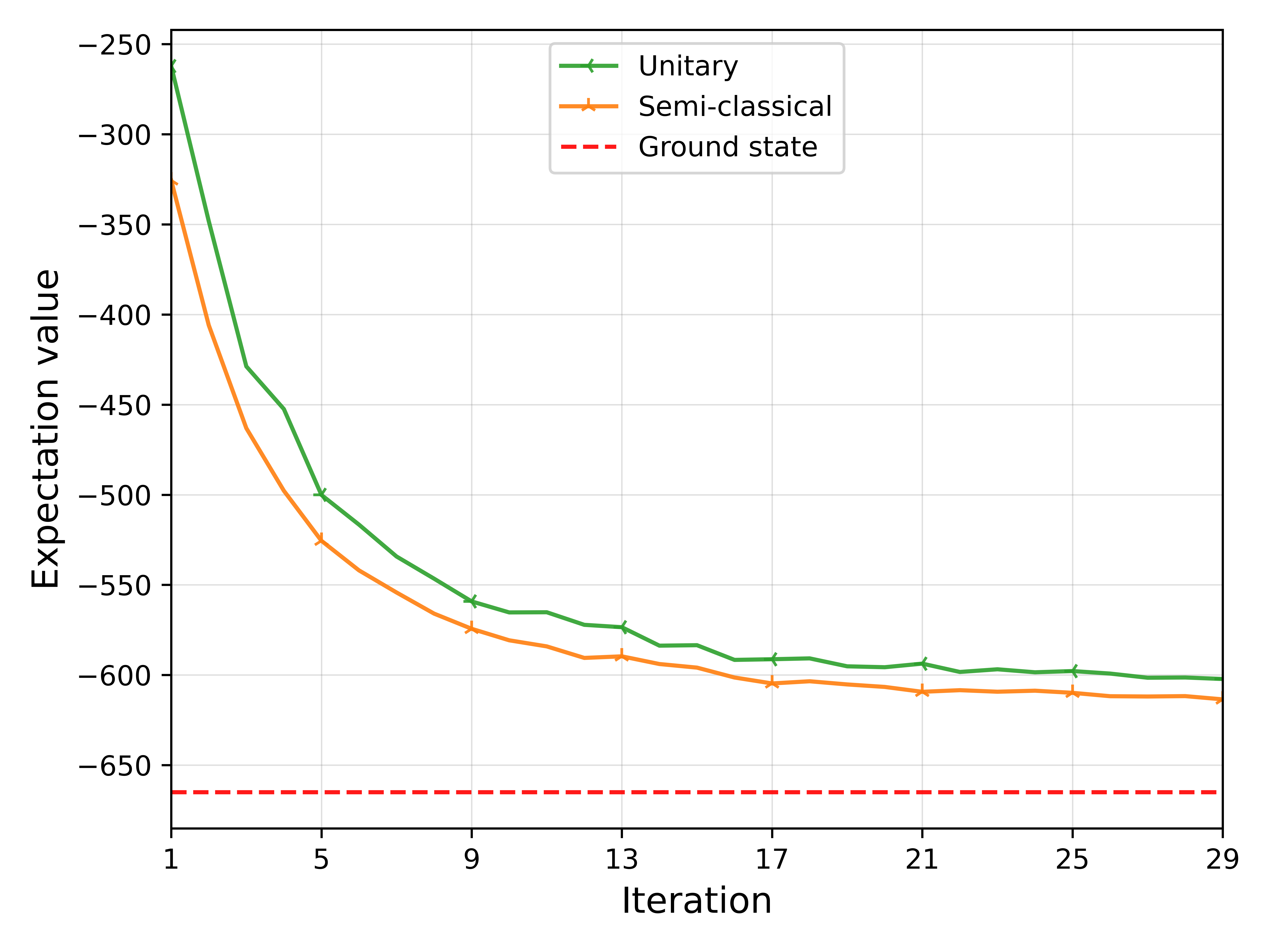}
    \caption{Expectation value of the problem Hamiltonian for a 30-qubit exact cover instance.
    We show results from a unitary implementation (with and without semi-classical simplification) using $M=30$ and $\Delta\tau=0.003$.
    The lines start at iteration $m=1$ of QITE, as all methods yield zero expectation value before the first iteration.
    The dashed red line indicates the ground-state energy.
    In the final state, the semi-classical implementation sampled one of four ground-state solutions exactly once out of $2^{14}$ shots, resulting in a negligible empirical success probability $\psuccess$.
    }
    \label{fig:hardware30qb}
\end{figure}

For the 30-qubit instance shown in Fig.~\ref{fig:hardware30qb} (\verb|ibm_pittsburgh|), the semi-classical approach dependency of the unitary circuits becomes apparent.
The gap between the methods did not have any striking expansion when increasing the qubit count.
Certain observables required for parameter updates, including $Y_kY_i$, $X_k$, $X_kZ_i$, and $X_kX_i$, cannot be implemented using a purely semi-classical fan-out.
In particular, the terms $Y_kY_i$, $X_k$, and $X_kZ_i$ contribute to the update of the $Y_k$ rotation parameter, which is essential for the success of the reduced ansatz.
As a result, the semi-classical implementation does not offer a clear advantage over the fully unitary or dynamic implementations in this regime.

\subsubsection{Dynamic circuit requirements}
\label{sec:dynamic_circuits_requirements}
Across all hardware experiments, the dynamic circuit implementation yielded larger errors than the unitary and semi-classical realizations, despite its constant entangling-gate depth.
IBM reports parallel execution of feedback operations provided they act on different qubits \cite{IBM_dynamiccircuits_blog_2025}.
In our constructed circuits, feedback acts on distinct qubits and uses separate classical registers, yet inspection of the scheduled circuits reveals non-parallel execution.
Consequently, while the entangling depth remains constant, the overall circuit depth scales linearly with system size due to serialized single-qubit feedback operations.
Solving the engineering issue of parallel feedback would yield better results with current hardware.
To quantify the ability of the dynamic circuits in the parallel feedback setting, we analyze the complete process fidelity rather than circuit depth alone.
The process fidelity accounts for idling errors, two-qubit gate errors, measurement errors, and classical feed-forward latency.

In Appendix~\ref{app:proccess_fidelity}, we compare lower bounds on the process fidelities of the unitary and dynamic implementations as a function of system size and hardware error parameters.
Using the reported error rates and operation times of the \verb|ibm_pittsburgh| backend, the fidelity lower bounds of the two approaches intersect at approximately $45$ qubits, at a fidelity of $\sim 0.15$.

Achieving a crossover at a higher, somewhat relevant fidelity of $\sim 0.5$ would require reductions of the median CNOT and measurement error rates to approximately $0.6\times10^{-3}$ and $1.4\times10^{-3}$, respectively, together with a feedback latency of $3\,\mu\mathrm{s}$.

\section{Conclusion}
\label{sec:conclusion}
In this work, we explored the ability to reduce demand on quantum hardware through dynamic circuits on exact cover and set partitioning instances for the QITE algorithm.
The suggested fan-out based approach necessitates the entanglement distribution of the circuit to be concentrated to a single, or at least a constant number of qubits.
Beyond the structural assumptions underlying the reduced ansatz, an important empirical observation of this work is that restricting the operator set does not degrade performance in noiseless simulations.
On the contrary, for the difficult exact cover and set partitioning instances considered here, the reduced-parameter construction consistently achieves higher success probabilities than the more expressive P2A variants, which utilizes all real-valued two-qubit rotations.
In addition, the compression of successive QITE layers, which introduces a controlled first-order approximation error, does not deteriorate performance and in several cases leads to improved convergence behavior allowing the method to outperform previous quantum attempts on the same instances \cite{Svensson_2023_Hybrid, cacao_2024_the}.
These findings indicate that, for the dense diagonal instances studied, increasing the expressivity of the operator set does not necessarily translate into improved ground-state convergence.

At present, we do not provide a formal theoretical explanation for this effect.
One possible interpretation is that, for the problem instances considered, the relevant low-energy manifold is effectively accessible within the restricted operator set, such that additional operators mainly increase the dimensionality of the parameter space without facilitating convergence.
From this perspective, the reduced ansatz may act as an implicit regularization of the QITE update rule.
However, establishing whether this interpretation holds more generally, and identifying structural properties that predict when such a reduction is beneficial, remains an open question.
Compared to previous simplifications of ITE-based methods \cite{Nishi_Kosugi_Matsushita_2021,Silva_2023, Gacon_Variational_2024}, ours is not at algorithmic level, but motivated by the structure of the problem instances \cite{buhrman2025formalframeworkquantumadvantage}.
Recognizing the improved success probability, it is vital to characterize what properties beyond density allow simplification, and what computational problems can elicit such properties.
Further combining the reduced ansatz with additional simplifications such as the removal of the Quantum Geometric Tensor for VarQITE \cite{Gacon_Variational_2024, Fitzek2024} is an interesting avenue for future work.

Regarding the hardware results, the experiments clearly demonstrate that constant entangling depth alone is not sufficient to guarantee improved performance on current devices.
Although the dynamic implementation realizes the intended interaction pattern with fixed two-qubit depth, the additional measurement operations, classical feed-forward, and associated noise sources currently outweigh the depth advantage, which is further reduced by the observed serialization of feed-forward operations.
On the other hand, the semi-classical implementation, which is a simple dynamic circuit, achieves the most stable performance on the 20 and 30 qubit instances.
The 20-qubit instance which it manages to solve is the largest application of QITE in real quantum hardware, to our knowledge.
Additionally, most implementations of dynamic circuit protocols have focused on specific subroutines or fundamental results~\cite{baumer_2024_efficient, baumer_measurement-based_2024,Iqbal_2024, bäumer2024quantumfouriertransformusing,song_constant_2025,kanno2025efficientimplementationrandomizedquantum, fossfeig2023experimentaldemonstrationadvantageadaptive, wu2025measurementandfeedbackdrivennonequilibriumphase, waring2025chshviolationsusingdynamic}, whereas end-to-end applications are not as prominent \cite{C_rcoles_2021, pokharel2025orderchaosadaptivecircuits} although suggestions exists \cite{tserkis2025depthoptimizationansatzcircuits, Plathanam_Babu_2025}.
Here we take another step towards practical use of dynamic circuits by implementing the previously suggested dynamic fan-out gate in the QITE algorithm.

We consider the fidelity of the QITE ansatz with respect to its noiseless counterpart to characterize the regime in which crossovers occur, making explicit the hardware improvements required for dynamic circuits to become advantageous in this setting.
Currently, circuit sizes with $\gtrsim 45$ system qubits allow higher fidelities with the dynamic approach (with $\gtrsim 89$ qubits), although at an impractically low final state fidelity.
Imposing a fidelity threshold of $0.5$, we identify that both error rates of measurements and two-qubit gates need to be reduced by approximately $65\%$, combined with twice as fast feedback operations. 
Under these conditions, the crossover shifts to circuits of around $29$ system qubits, that is, $57$ qubits in the dynamical implementation.
Thus, rather than providing a negative result, the experiments delineate concrete performance thresholds for future devices.
In general, improved QITE state fidelity leads to an increased ground‑state success probability $\psuccess$; however, a detailed analysis across problem instances is required to characterize the implementations' performance with respect to the classical problems. 
Table~\ref{tab:hardware_results_pgs} shows the success probabilities $\psuccess$ for different implementations, with dynamic circuits showing an average value of $0.004$ for 14 qubit instances.

Taken together, these results suggest that algorithmic simplification and hardware-aware circuit design must be considered jointly.
The reduced ansatz enables depth compression and dynamic implementations, but the practical benefit depends sensitively on the noise characteristics and control latencies of the target platform.
Further left to investigate is what other quantum algorithms could benefit from similar diminution, both performance and hardware-wise.
 \let\oldaddcontentsline\addcontentsline
\renewcommand{\addcontentsline}[3]{}

\begin{acknowledgments}
We thank Göran Johansson, Martin Andersson and Mattias Grönkvist for valuable insights and fruitful discussions. 
A. L., E. M. and L. G.-\'{A}. acknowledge support from the  Knut and Alice Wallenberg Foundation through the Wallenberg  Center for Quantum Technology (WACQT).
W. D. acknowledges funding from the German Federal Ministry of Research, Technology and Space (BMFTR) under the research program Quantensysteme and funding measure Quantum Futur 3 for project No. 13N17229 as well as the Helmholtz Association via Initiative and Networking Fund for projects No. VH-NG-21-08 under the Helmholtz Investigator program as well Projects KA-QUS-02 (qFLOW) and KA-QUS-03 (QT-Batt) under the Helmholtz Quantum Use case call. 
\end{acknowledgments}

\bibliography{references}

\appendix

\section{Heuristic for choosing $k$ in the reduced parameter ansatz}
\label{sec:app_heur}
A simple heuristic for choosing $k$ when the optimal choice is not known is to select the variable with the lowest total weight induced by the Hamiltonian, \textit{i.e.}\ the minimizer of
\begin{equation}
    w_k = h_k + \sum_{r' \in R} J_{kr'} .
\end{equation}
Intuitively, this favors variables whose associated subsets cover many elements, rather than variables that mainly contribute additional overlaps with other subsets.
In the worst case, identifying a suitable $k$ incurs a linear overhead in the number of restarts, which may or may not be acceptable depending on the total runtime.

\begin{figure}[htbp]
    \centering
    \includegraphics[width=1\linewidth]{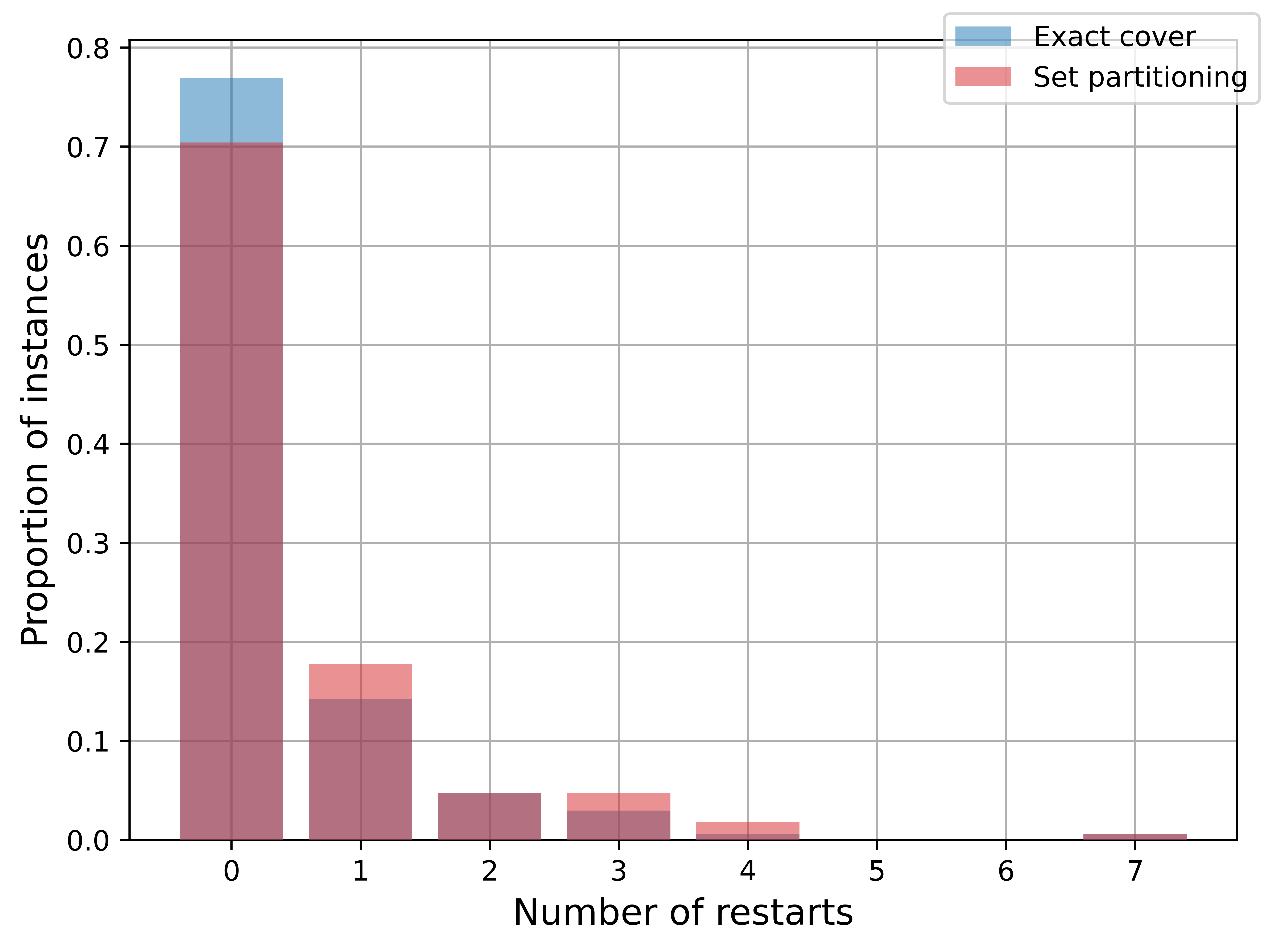}
    \caption{Distribution of suitable $k$ obtained from the heuristic for instances of size 6--120.
    For exact cover, ``$\sim 75\%$ at 0 restarts'' means that for $\sim 75\%$ of the instances the smallest $w_k$ corresponds to a variable that appears in at least one ground state, resulting in zero overhead.
    For $\sim 15\%$ of the instances, the second smallest $w_k$ yields a ground-state variable, requiring one restart, and so on.}
    \label{fig:wr_distribution}
\end{figure}

Figure~\ref{fig:wr_distribution} shows the distribution of suitable choices of $k$ obtained by this heuristic across all instances, both with and without the objective function.

\section{Linear QITE performance}
\label{app:linear_QITE_performance}
\begin{figure}[htbp]
    \centering
    \includegraphics[width=1\linewidth]{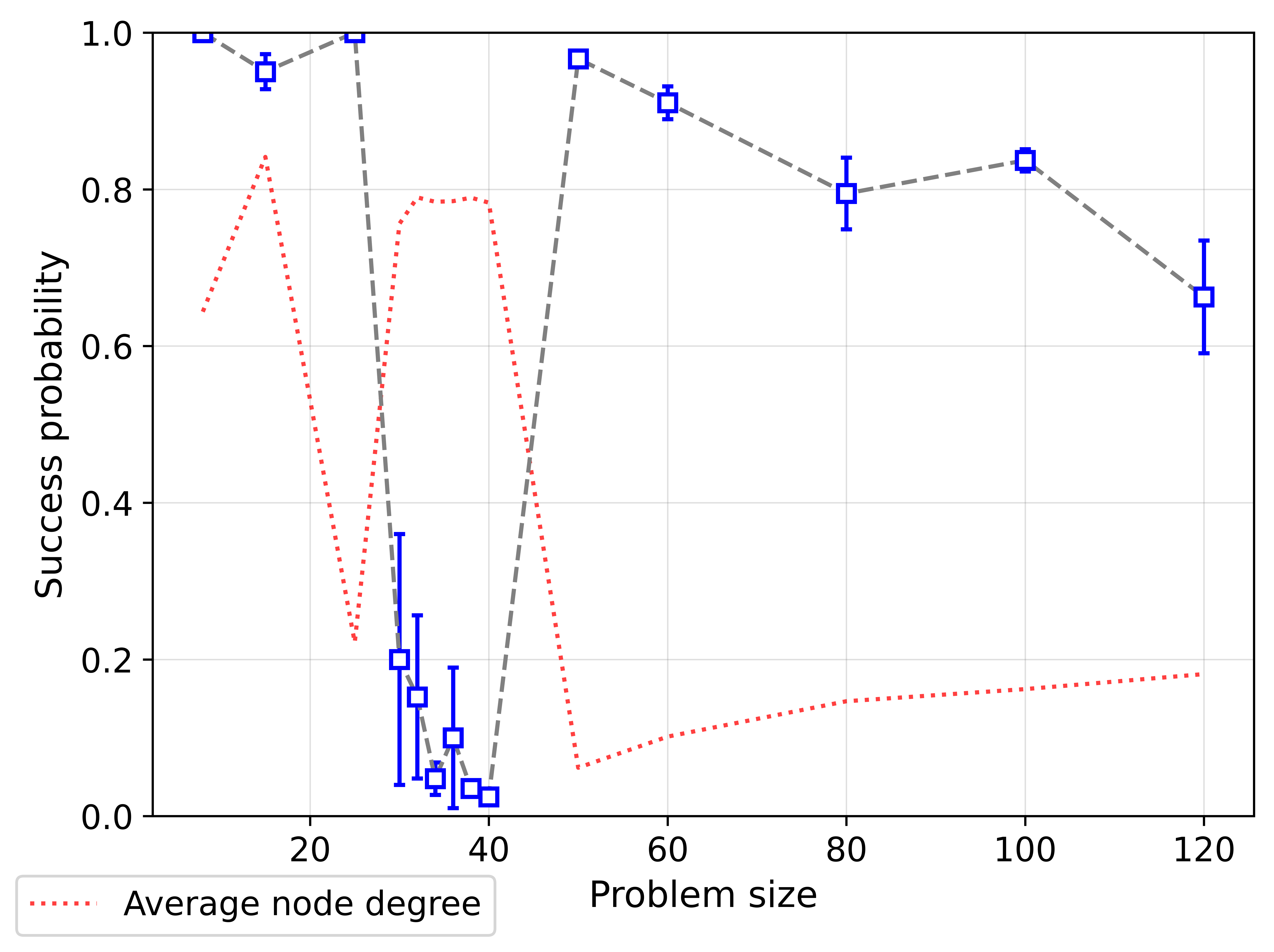}
    \caption{Success probability $\psuccess$ of the linear ansatz for different problem sizes using $M = 5000$ iterations and $\Delta\tau = 0.0016$.
    The large number of iterations is justified by their polynomial cost, $O(N^2)$.
    The error bars show the variance $\sigma^2$.
    The dashed red line represents the average node degree of the instances divided by the maximum node degree, $N-1$, indicating the proportion of other subsets with which each subset shares at least one element on average.}
    \label{fig:exact_cover_linear_ansatz}
\end{figure}

As noted by \textcite{Vikstal_2020_applying}, the most difficult instances for QAOA are those with dense conflict graphs.
Since linear QITE corresponds to gradient-descent VQE \cite{xie2025adaptiveweightedqitevqealgorithm}, one expects dense problem instances to be particularly challenging for linear QITE as well.
The results for the linear ansatz are presented in Fig. \ref{fig:exact_cover_linear_ansatz}.

\begin{figure}[htbp]
    \centering
    \includegraphics[width=1\linewidth]{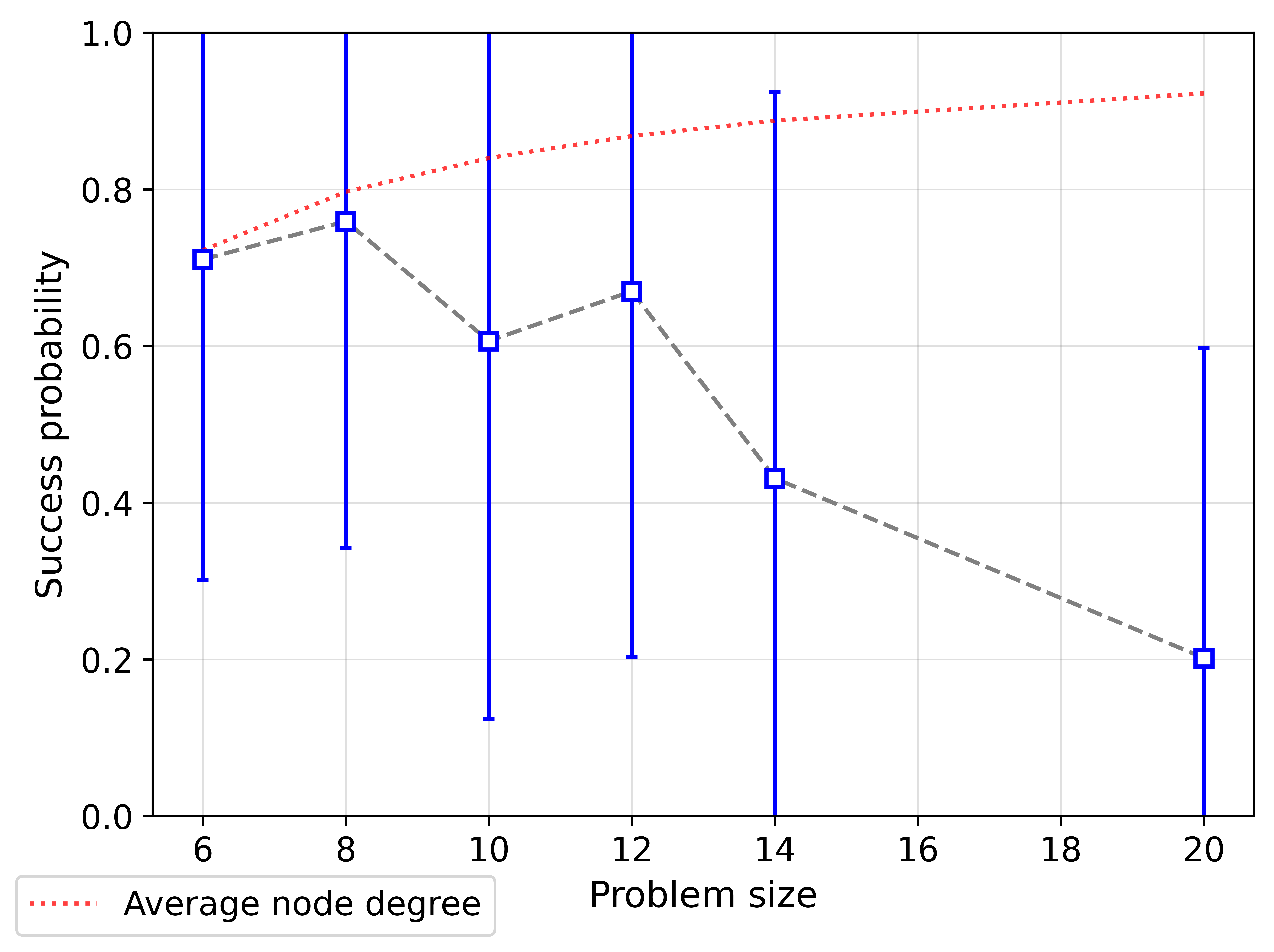}
    \caption{Success probability $\psuccess$ of the dense instances from Ref.~\cite{Svensson_2023_Hybrid}.
    The large variance $\sigma^2$ reflects the method's tendency either to converge to a ground state or to yield negligible overlap.
    In this regime, the success probability effectively corresponds to the fraction of instances that converge to the ground state.}
    \label{fig:exact_cover_linear_ansatz_dense}
\end{figure}

As expected, conflict density has a strong impact on performance, although the problem size also appears to play a role.
Since the dense instances with poor performance exceed the system sizes accessible to simulations of higher-order ansätze, we additionally consider smaller instances with higher node degree taken from \cite{Svensson_2023_Hybrid}.
In Fig.~\ref{fig:exact_cover_linear_ansatz_dense}, we observe a substantial decrease in performance, further illustrating the limitations of linear QITE.

\begin{figure}[htbp]
    \centering
    \includegraphics[width=1\linewidth]{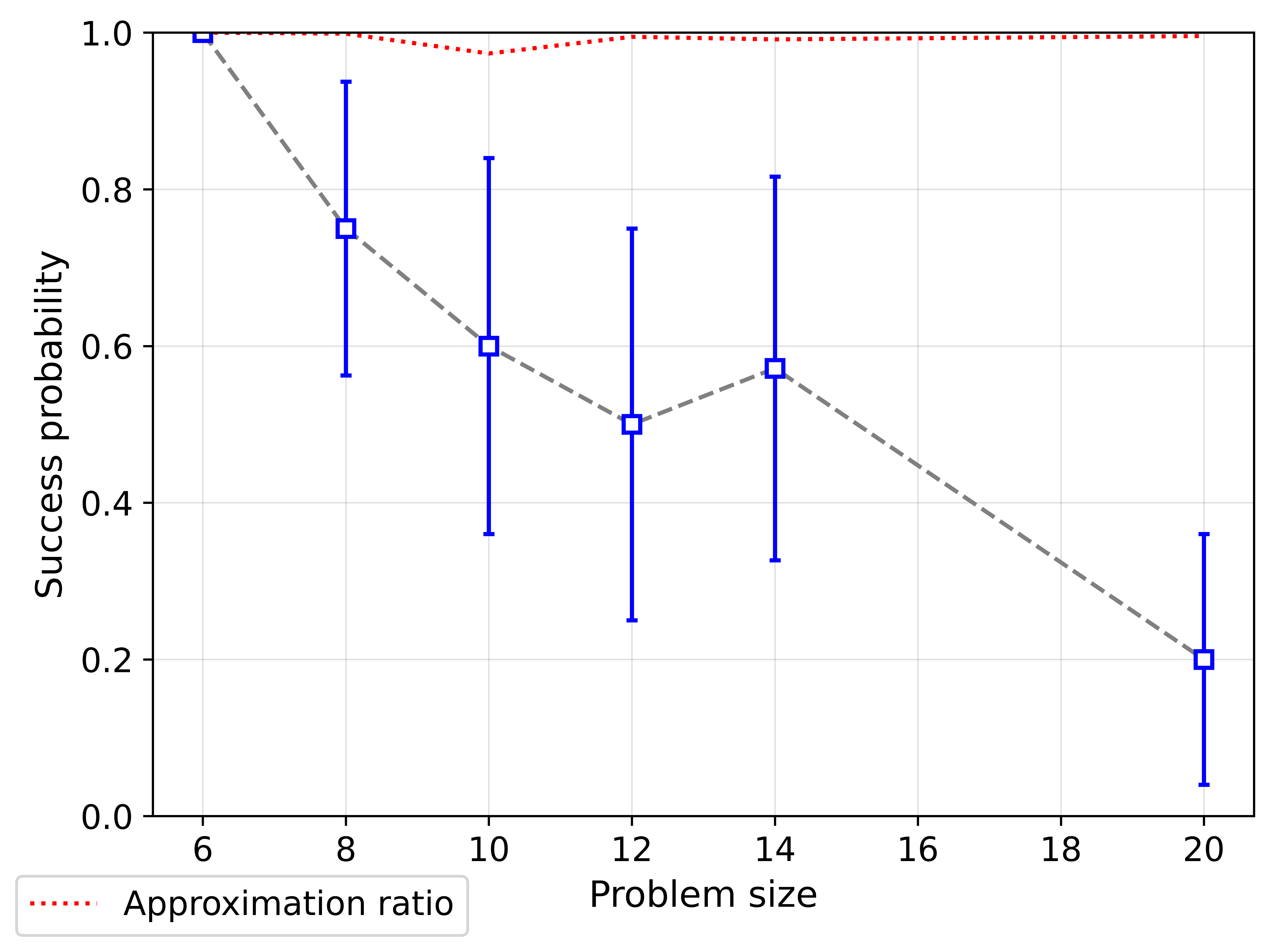}
    \caption{Success probability of the linear QITE algorithm on the set partitioning instances using $M = 5000$ and $\Delta\tau = 2.5\times10^{-5}$.
    The dotted red line shows the approximation ratio, illustrating the efficiency of linear QITE in finding low-energy states.}
    \label{fig:linear_qite_set_partitioning}
\end{figure}

Finally, Fig.~\ref{fig:linear_qite_set_partitioning} shows the performance of linear QITE on the set partitioning instances.

\section{Circuit compression}
\label{app:compression}
As shown by \textcite{Nishi_Kosugi_Matsushita_2021}, if the $(N_{\text{comp}}+1)$-th step is compressed into the $N_{\text{comp}}$-th step, the leading error term is
\begin{equation}
    \frac{1}{2}\Delta\tau^2 [\Aop{N_{\text{comp}+1}}],\sum_{m=1}^{N_{\text{comp}}}\Aop{m}] + \mathcal{O}(\Delta\tau^3).
\end{equation}
For a single pivot qubit $k$ with
\begin{equation}
    \Aop{N_{\text{comp}+1}} = \sum_i \acoeff{N_{\text{comp}+1}}{i}Y_i + \sum_{i\neq k}\bcoeff{N_{\text{comp}+1}}{i}Z_kY_i   
\end{equation}
and
\begin{equation}
    \sum_{m=1}^{N_\text{comp}}\Aop{m} = \sum_i \sum_{m=1}^{N_\text{comp}}\acoeff{m}{i} Y_i + \sum_{i\neq k} \sum_{m=1}^{N_\text{comp}}\bcoeff{m}{i} Z_kY_i,
\end{equation}
one obtains
\begin{align}
    &\left[\Aop{N_{\text{comp}+1}}, \sum_{m=1}^{N_{\text{comp}}}\Aop{m}\right] = \sum_{i \neq k} \Big( \acoeff{{N_{\text{comp}+1}}}{k} \nonumber \\
    & \qquad \times \sum_{m=1}^{N_\text{comp}}\bcoeff{m}{i} -\bcoeff{{N_{\text{comp}+1}}}{i}\sum_{m=1}^{N_\text{comp}} \acoeff{m}{k} \Big)X_kY_i.
\end{align}

Higher order errors will precisely correspond to the DLA of the generators, due to the Baker-Campbell-Hausdorff formula.
After compression, the ansatz is no longer invariant under the removal of $XY$ rotations, or more generally operators containing an odd number of $Y$ terms.
Instead, compression further restricts the set of reachable states by reducing the effective parameter space, and consequently also limits the spread of entanglement.

With increasing iterations, any slowdown in parameter updates is therefore expected to arise from the algorithm approaching an eigenstate rather than from circuit-depth limitations, which remain constant.
In this sense, compression does not introduce an additional depth-induced bottleneck and may even improve practical performance.

Including additional pivot qubits leads to further second-order correction terms, for instance $[Z_kY_i,Z_jY_k] \propto X_kY_iZ_j$.
These additional contributions imply that each compression step introduces larger higher-order errors, including not only $XY$ terms but also three-qubit interactions.

\subsection{Choice of two-qubit rotations}
Finally, one must choose whether to use $ZY$ or $XY$ rotations.
Following \textcite{Zanardi_Zalka_Faoro_2000}, the entangling power $e_p$ of a two-qubit gate $U$ with respect to an entanglement measure $E$, evaluated on a product state $\ket{\psi} = \ket{\psi_1} \otimes \ket{\psi_2}$, is defined as
\begin{equation*}
    e_p(U) := \overline{E(U\ket{\psi_1}\otimes \ket{\psi_2})}^{\psi_1,\psi_2}, 
\end{equation*}
where the bar denotes an average over a chosen distribution of product states.
We use the linear entropy,
\begin{equation*}
    E(\ket{\Psi}) := 1-\text{Tr}(\rho^2), \quad \rho := \text{tr}_2 \ket{\Psi}\bra{\Psi}
\end{equation*}
which is convenient since it is a polynomial in  $\ket{\psi}$.

Because $R_{XY}(\theta)$ and $R_{ZY}(\theta)$ are locally equivalent,
\[
R_{X_iY_j}(\theta) = H_i R_{Z_iY_j}(\theta) H_i,
\]
with $H_i$ the Hadamard gate on qubit $i$, their average entangling power is identical \cite{Morachis_Galindo_Maytorena_2022}.
However, when restricting to product states of the form
\[
\ket{\psi_1} = a\ket{0} + b\ket{1}, 
\qquad
\ket{\psi_2} = c\ket{0} + d\ket{1},
\]
with real, positive amplitudes $0 \le a,b,c,d \le 1$, as motivated by the problem instances considered here, the entangling power of $XY$ and $ZY$ differs.

\begin{figure}[H]
    \centering

    \makebox[0.45\textwidth][l]{(a)}%
    \par\vspace{-1.2em}
    \begin{minipage}{0.45\textwidth}
        \centering
        \includegraphics[width=1\textwidth]{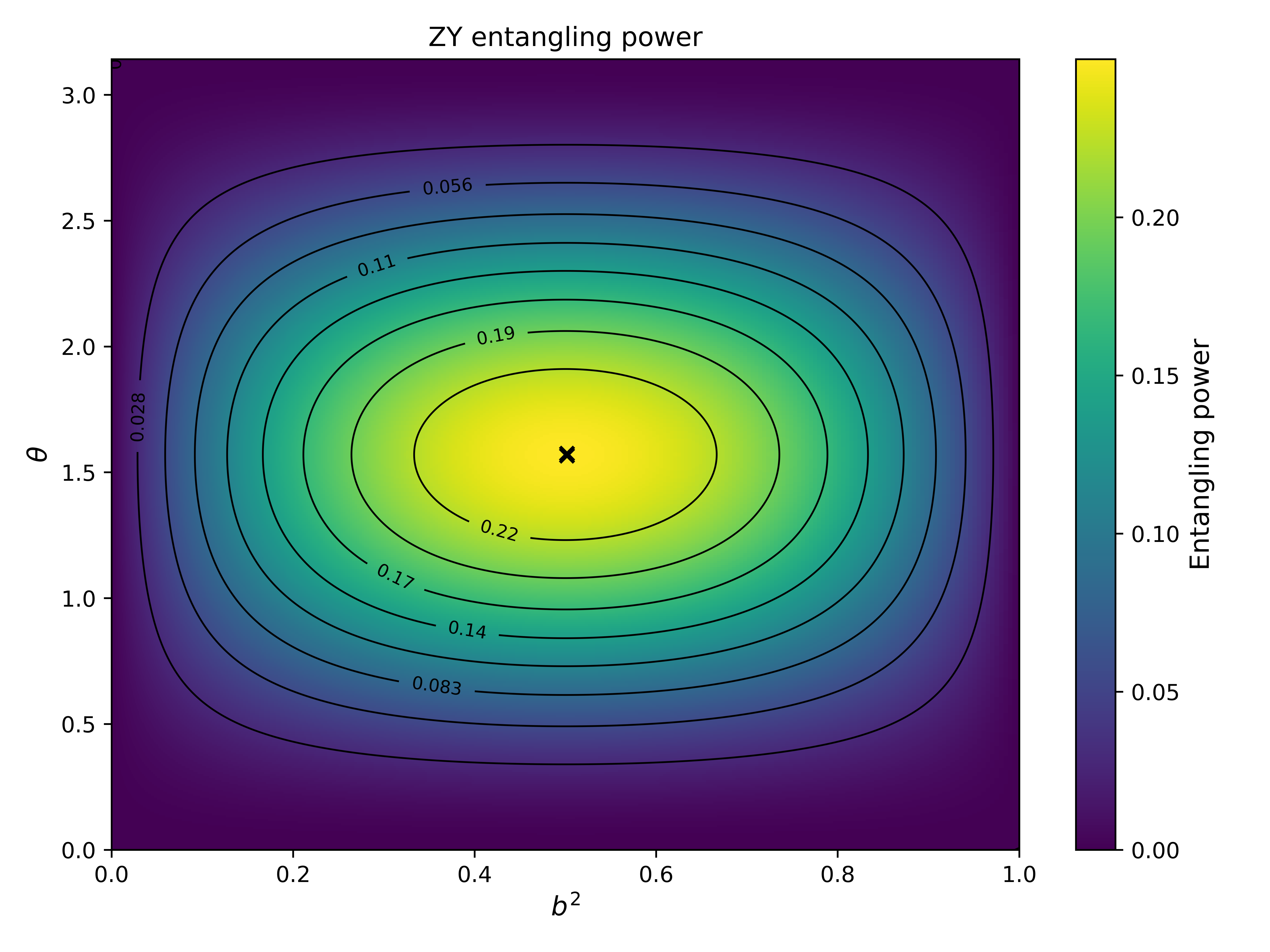}
    \end{minipage}
    \hspace{1em}
    \makebox[0.45\textwidth][l]{(b)}%
    \par\vspace{-1.2em}
    \begin{minipage}{0.45\textwidth}
        \centering
        \includegraphics[width=1\textwidth]{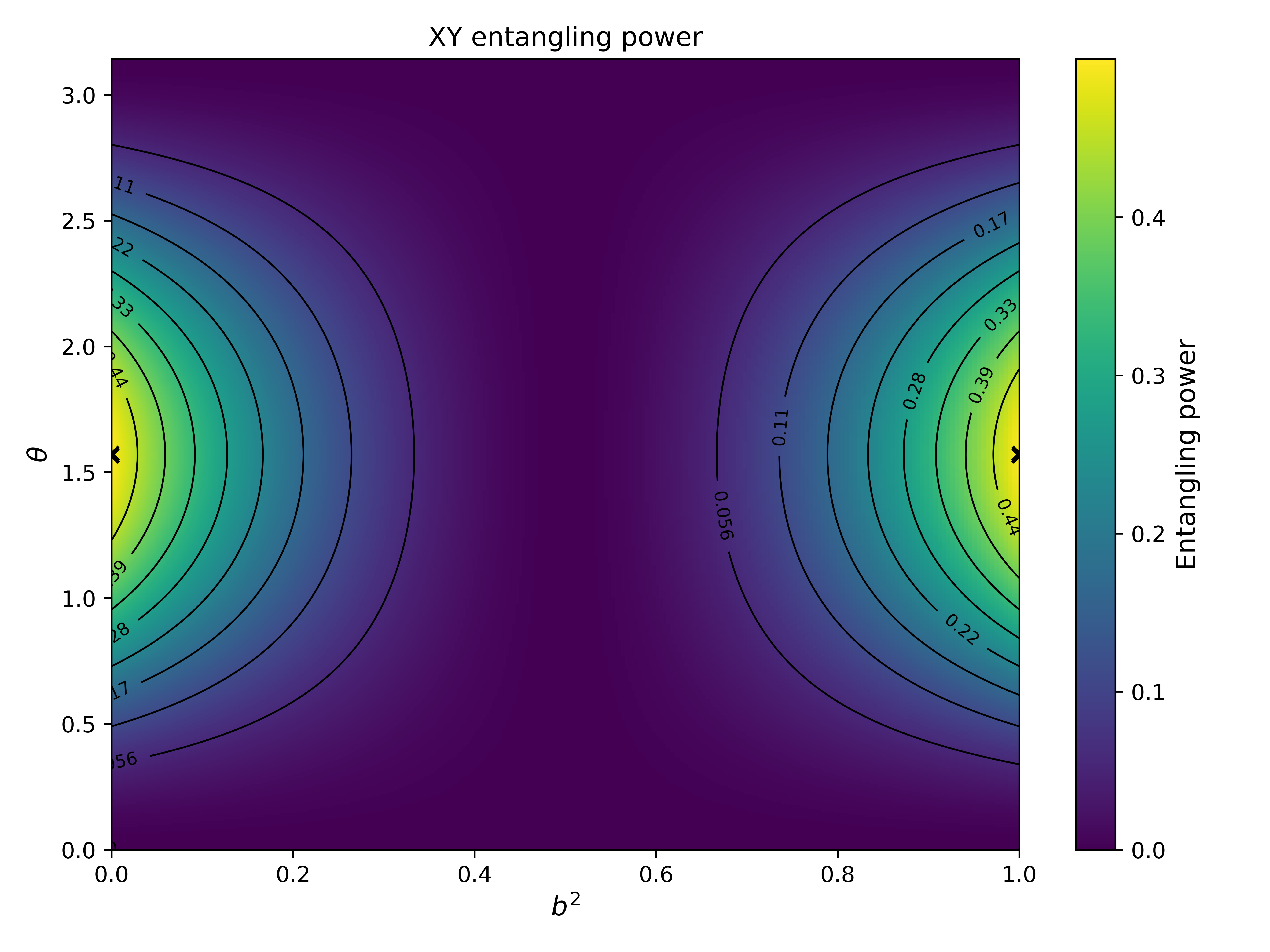}
    \end{minipage}

    \caption{
    (a) Entangling power of an $R(\theta)_{Z_iY_j}$ rotation as a function of the angle $\theta$ and the probability of measuring qubit $i$ in state $\ket{1}$.
    The maximum is marked by a cross at $(1/2,\pi/2)$.
    (b) Same as (a) but for an $R(\theta)_{X_iY_j}$ rotation.
    The maximum occurs at $(0,\pi/2)$ and $(1,\pi/2)$.
    }
    \label{fig:entangling_power}
\end{figure}

Specifically,
\begin{eqnarray}
    \label{eq:entangling_power}
    &E(e^{-i\frac{\theta}{2}XY}\ket{\psi_1}\otimes \ket{\psi_2}) = \frac{1}{2}(1-2b^2)^2\sin^2\theta,\nonumber\\
    &E(e^{-i\frac{\theta}{2}ZY}\ket{\psi_1},\otimes \ket{\psi_2}) = b^2(1-b^2)\sin^2\theta. \nonumber
\end{eqnarray}
In this restricted setting, the entangling power depends only on the amplitude $b$.
Averaging over $b \in [0,1]$ yields
\begin{align*}
    &\int_0^1 \frac{1}{2}(1-2b^2)^2\sin^2\theta db = \frac{7}{30}\sin^2\theta,\\
    &\int_0^1 b^2(1-b^2)\sin^2\theta db = \frac{8}{30}\sin^2\theta.
\end{align*}
Thus, within this class of states, $R_{ZY}$ rotations generate slightly more entanglement on average than $R_{XY}$ rotations.
Both attain their maximal entangling power at $\theta = \pm \frac{\pi}{2}$.
Figure~\ref{fig:entangling_power} further illustrates that for $R_{ZY}$ rotations the maximum occurs at $b = 1/2$, whereas for $R_{XY}$ rotations it occurs at $b \in \{0,1\}$.

Since the algorithm starts in the state $\ket{+}^{\otimes N}$, $R_{XY}$ would leave it in a product state, whereas this is precisely the point at which $R_{ZY}$ exhibits maximal entangling power.
Moreover, if the final state is close to a product state and the angles satisfy $\theta_{Y_i} \approx \pm \frac{\pi}{2}$, no additional entanglement is introduced, consistent with convergence.
The preference for $ZY$ over $XY$ rotations is also reflected in Fig.~\ref{fig:exact_cover_p2a_ansatz} where ansätze utilizing only $ZY$ rotations outperform those with only $XY$.

\section{Experimental details}
\label{sec:exp_details}
The hardware experiments for the 20- and 30-qubit instances were carried out on the \verb|ibm_pittsburgh| backend.
The remaining experiments used \verb|ibm_boston|.

Since the algorithm starts in the state $\ket{+}^{\otimes N}$ with all parameters initialized to zero, the first QITE layer can be computed classically in a straightforward manner.
For this initial state, any Pauli string containing a $Z$ or $Y$ operator has vanishing expectation value.
Consequently, the linear system in the first iteration reduces to
\begin{align*}
    a_{i} &=\Delta\tau h_i \quad \forall i,\\
    b_{i}
    &= \Delta\tau J_{ki} \quad \forall i.
\end{align*}

On hardware, expectation values were measured by grouping observables according to qubit-wise commutativity (QWC) \cite{Kandala_2017,Verteletskyi_2020}.
All experiments employed dynamical decoupling as an error-mitigation technique.
Expectation values were estimated using $2^{14}$ shots.

The semi-classical implementation of the ansatz is applicable to groups in which the pivot qubit $k$ is measured in the $Z$ basis.
To maximize the benefit of the semi-classical approach, the grouping was arranged such that only a minimal number of observables remained in groups that could not exploit the semi-classical fan-out.
With QWC grouping, there were at most two such groups: one measuring $X$ on qubit $k$ and one measuring $Y$ on qubit $k$.
These groups were implemented unitarily.

The dynamic circuits made use of the dedicated ``Mid-circuit measure'' instruction, which has a shorter duration than standard measurement operations.
Reset of the ancilla qubits was implemented via feed-forward, immediately following their measurement.

The unitary implementation employed a SWAP network, resulting in a depth that scales linearly with system size.
An ASAP scheduler was used for the unitary circuits to measure qubits as early as possible, whereas the dynamic and semi-classical implementations employed an ALAP scheduler.

\subsection{Post-processing}
In principle, both the circuit in Fig.~\ref{fig:quantum_circuits} and the one in Fig.~\ref{fig:semi-classical_circuit} could be realized via post-selection instead of explicit feed-forward operations.
Previous experiments demonstrating the advantage of dynamic implementations of a single quantum fan-out gate have relied on post-processing to extrapolate the scaling to larger systems \cite{baumer_measurement-based_2024, song_constant_2025}.

For the circuits considered here, the final corrections associated with the last fan-out gate in Fig.~\ref{fig:quantum_circuits} could be performed in post-processing.
However, extending this approach to the entire ansatz is considerably more challenging.
In particular, when non-Clifford operations follow the fan-out gate, post-processing no longer provides an error-free emulation of feed-forward \cite{ruh2024quantumcircuitoptimisationmbqc}.

Moreover, pure post-selection would introduce an exponential sampling overhead due to the multiple feed-forward operations that depend on XOR combinations of measurement outcomes.
Post-selection fixes a specific pattern of measurement-dependent corrections and discards all other outcomes.
For example, enforcing identity corrections would require the first ancilla to measure 0 and all subsequent relevant auxiliary qubits to measure 0 as well.
Using $N-1$ additional qubits, with $(N-1)/2$ of them contributing to $X$ corrections, only one out of the $2^{(N-1)/2}$ possible measurement outcomes would be retained under post-selection.
The same reasoning applies to any other fixed choice of correction pattern.
Without additional assumptions about the quantum state, the required sampling overhead therefore scales exponentially.
For this reason, the dynamic and semi-classical implementations were realized using hardware feed-forward operations rather than post-selection.

\section{Process fidelity}
\label{app:proccess_fidelity}
Following the method of Ref.~\cite{baumer_2024_efficient}, the final process fidelity of a circuit can be loosely lower bounded by $e^{-\lambda_{\text{tot}}}$, where $\lambda_{\text{tot}}$ is a single Pauli noise rate given as a sum of independent decoherence contributions:
\begin{equation}
    \lambda_{\text{tot}} = t_{\text{idle}}\lambda_{\text{idle}} + N_{\text{CNOT}}\lambda_{\text{CNOT}} + N_{\text{meas}}\lambda_{\text{meas}}.
\end{equation}
The idle time $t_{\text{idle}}$, the number of CNOT gates $N_{\text{CNOT}}$, and the number of measurements $N_{\text{meas}}$ depend on the implementation (unitary versus dynamic), as summarized in Table~\ref{tab:method_lambda}.
The remaining parameters are hardware dependent.
    
\begin{table}[h]
    \centering
    \caption{\label{tab:method_lambda} Idle time, number of CNOT gates and measurements for the different implementations.
    $j= {t_\text{measure}}/{3t_\text{CNOT}}$, \textit{i.e.} the number of target qubits which begin to idle, waiting for previous measurements to finish.
    $\mu_\text{Dynamic}$ and $\mu_\text{SC}$ represent the idle time incurred by measurement and feedback operations expressed as multiples of the CNOT gate time.
        $\mu_{\text{Dynamic}} = (2t_\text{measure}+3t_\text{feedback})/t_\text{CNOT}$ and $\mu_\text{SC} = (t_\text{measure}+2t_\text{feedback})/t_\text{CNOT}$.
    The extra feedback operation in $\mu_{\text{Dynamic}}$ comes from the requirement of a reset of the auxiliary qubits.}
    \begin{ruledtabular}
    \begin{tabular}{>{\bfseries}l c c c }
        \textbf{Implementation} & $t_{\text{idle}}$ & $N_{\text{CNOT}}$ & $N_{\text{meas}}$  \\
        \midrule
        Unitary & $\frac{3}{2}(N-1)(N+j-1)$ & $3N-4$ & 0\\
        Dynamic & $4 + 2N + N\mu_{\text{Dynamic}}$ & $6N-8$ & $2N-2$ \\
        Semi-classical & $(N-1)\mu_\text{SC}$ & 0 & 1 \\
    \end{tabular}
    \end{ruledtabular}
\end{table}

The idle time for the unitary implementation reflects current hardware constraints, which ideally would not be present.
Although the unitary circuits have linear depth, only the control qubit must remain active throughout the entire circuit.
In principle, the remaining qubits could be measured immediately after interacting with the control qubit and initialized only when needed.
In practice, however, sequential initialization induces cross-talk errors, and qubits sharing the same readout line can only be measured in parallel if measurement starts simultaneously.
As a result, some qubits must wait before measurement, leading to idle-time errors that increase with system size.

As discussed in Sec.~\ref{sec:dynamic_circuits_requirements}, the feedback operations were not performed in parallel for the executed circuits, although promised by IBM.
Depending on the latency of classical processing, a practical advantage may still arise.
For the fidelity analysis, however, feedback operations are assumed to execute in parallel.

\begin{figure}[htbp]
    \centering
    \includegraphics[width=1\linewidth]{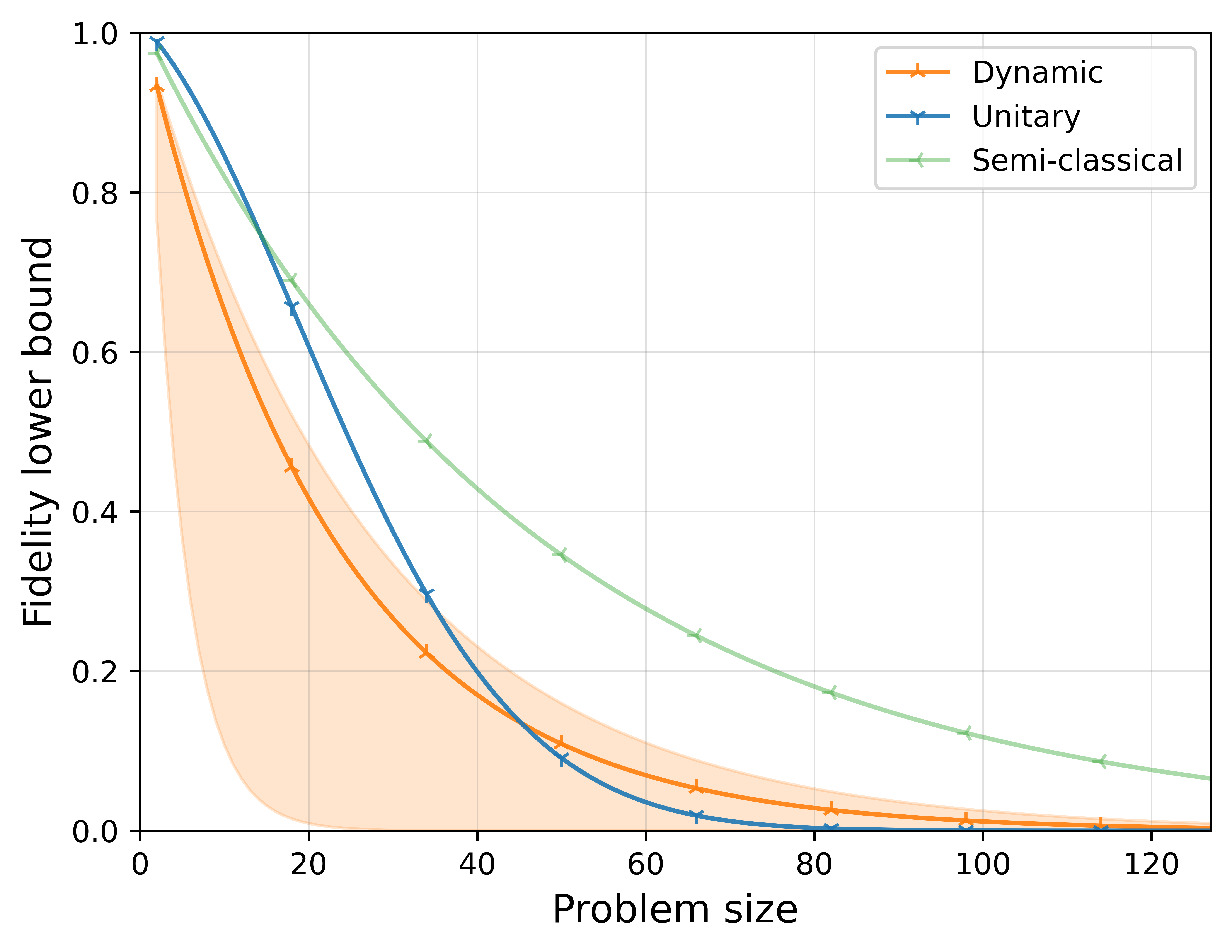}
    \caption{Lower bound on the fidelities of the different implementations.
    The shaded region corresponds to the dynamic approach using the worst- and best-case reported measurement errors.
    The line plot uses the median values.}
    \label{fig:lower_bound_fidelities}
\end{figure}

Using the reported median calibration errors from \verb|ibm_pittsburgh| for the hardware-dependent parameters, the resulting lower bound on the fidelity as a function of system size is shown in Fig.~\ref{fig:lower_bound_fidelities}.

\begin{figure}[H]
    \centering
    \includegraphics[width=1\linewidth]{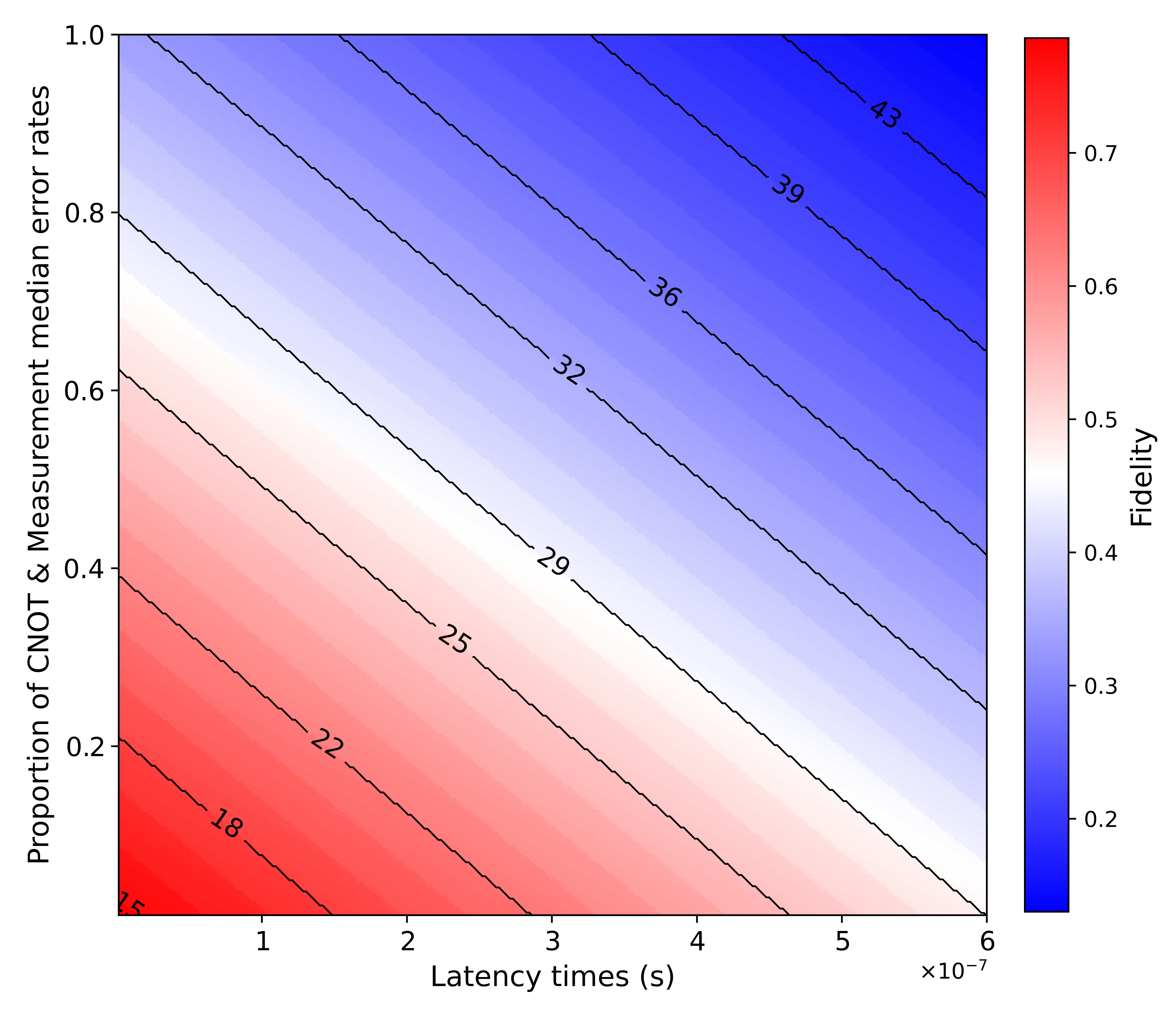}
    \caption{Heatmap of the lower bound fidelity and crossover point between unitary and dynamic implementations as a function of CNOT error, measurement error, and feedback latency.
    The y-axis shows the proportion of the currently reported IBM error rates (e.g., at $0.5$, errors are reduced by half).
    Contour lines indicate the system size at which the lower bound fidelity of the dynamic implementation exceeds that of the unitary one.
    The color scale gives the fidelity at the crossover point.
    Even eliminating CNOT and measurement errors alone is insufficient to achieve a practically relevant crossover; reductions in feedback latency are also required.}
    \label{fig:fidelity_crossover_heatmap}
\end{figure}

The fidelities do not cross until approximately 45 qubits, at which point the fidelity is already too low to be practically useful.
As seen in Table~\ref{tab:method_lambda}, the dynamic implementation involves more CNOT gates, more measurements, and additional feed-forward operations than the unitary implementation.
Reductions in the error rates of these operations, as well as shorter feedback latencies, would therefore benefit the dynamic approach most strongly.
Figure~\ref{fig:fidelity_crossover_heatmap} illustrates the required reductions in error rates and latency for the dynamic implementation to become advantageous.\\

\end{document}